%% file: main.tex
\documentclass[a4paper,11pt]{article}

\pdfoutput=1
\usepackage{amsmath,amssymb,cite,tabularx,hyperref,graphicx,color,cancel,xspace,flafter,ifpdf,tablefootnote,todonotes,subcaption,float}

\newcommand{\HEJ}{{\tt HEJ}\xspace}
\newcommand{\HIGHEJ}{\emph{High Energy Jets}\xspace}

\def\spa#1.#2{\left\langle#1\,#2\right\rangle}
\def\spb#1.#2{\left[#1\,#2\right]}
\def\spaa#1.#2.#3{\langle\mskip-1mu{#1}
                  | #2 | {#3}\mskip-1mu\rangle}
\def\spbb#1.#2.#3{[\mskip-1mu{#1}
                  | #2 | {#3}\mskip-1mu]}
\def\spab#1.#2.#3{\langle\mskip-1mu{#1}
                  | #2 | {#3}\mskip-1mu\rangle}
\def\spba#1.#2.#3{\langle\mskip-1mu{#1}^+
                  | #2 | {#3}^+\mskip-1mu\rangle}
\def\spav#1.#2.#3{\|\mskip-1mu{#1}
                  | #2 | {#3}\mskip-1mu\|^2}
\def\jc#1.#2.#3{j^{#1}_{#2#3}}

\hypersetup{%
  colorlinks=true,
  linkcolor=blue,
  citecolor=blue
}

\title{High Energy Resummed Predictions for the Production of a Higgs Boson with at least One Jet}
\author{Jeppe~R.~Andersen$^{a}$, Hitham Hassan$^{a}$, Andreas Maier$^{b}$,
  J\'er\'emy Paltrinieri$^{c}$,\\ Andreas Papaefstathiou$^{d}$ and
  Jennifer~M.~Smillie$^{c}$\\\mbox{}\\
  $^a$ Institute for Particle Physics Phenomenology,\\University of Durham,
  South Road, Durham DH1 3LE, UK\\
  $^b$  Deutsches Elektronen-Synchrotron DESY, \\
  Platanenallee 6, 15738 Zeuthen, Germany\\
  $^c$ Higgs Centre for Theoretical Physics, University of Edinburgh,\\
  Peter Guthrie Tait Road, Edinburgh EH9 3FD, UK,\\
  $^d$Department of Physics, Kennesaw State University, Kennesaw, GA 30144, USA.}

\begin{document}
\begin{flushright}
DESY-22-164, IPPP/22/71, DCPT/22/142
\end{flushright}
\vspace{-1cm}
{\let\newpage\relax\maketitle}
\begin{abstract}
  We present all-order predictions for Higgs boson production plus at least
  one jet which are accurate to leading logarithm in $\hat s/|p_\perp |^2$.  Our
  calculation includes full top and bottom quark mass dependence at all orders in the
  logarithmic part, and to highest available order in the tree-level matching.
The
  calculation is implemented in the framework of High Energy Jets (HEJ).
  This is the first cross section calculated with $\log(\hat s)$ resummation and
  matched to fixed order
for a process requiring just one jet, and our results also extend the region
  of resummation for processes with two jets or more.
This is possible because the resummation is performed explicitly in phase space.
  We compare the results of our new calculation to
  LHC data and to next-to-leading order predictions and find a numerically
  significant impact of the logarithmic corrections in the shape of key
  distributions, which remains after normalisation of the cross section.
\end{abstract}

\tableofcontents
\input{Introduction}
\input{Theory}
\input{ComparisonData}

\input{Conclusions}
\appendix
\input{Appendix}

\input{QCDPlots}

\bibliographystyle{JHEP}
\bibliography{papers}

\end{document}

%% file: Introduction.tex
\section{Introduction}
\label{sec:introduction}

Analysing the Higgs sector is among the foremost objectives of the
LHC. To this end, experiments aim for accurate measurements of
processes where Higgs bosons are produced, either inclusively or in
association with other identified particles.
Given the phenomenological importance of processes involving Higgs
boson production, there are considerable efforts to provide
high-precision theory predictions. Perturbative corrections are
typically found to be sizable, necessitating the inclusion of effects
at high orders. This endeavour faces a major challenge: in large
regions of phase space Higgs boson production is predominantly
loop-induced, namely through gluon fusion via a virtual top-quark
loop. Inclusive gluon-fusion Higgs boson production
with full finite top-mass contributions
is currently known
at next-to-next-to-leading order (NNLO)~\cite{Czakon:2021yub}, exclusive Higgs
boson plus jet production at next-to-leading order (NLO)~\cite{Jones:2018hbb,Chen:2021azt},
and Higgs boson plus dijet production only at leading order
(LO)~\cite{DelDuca:2001eu,DelDuca:2001fn}.

To facilitate calculations, the top-quark mass is often assumed to be
much larger than all other scales. Based on this approximation, one
more order has been computed in the perturbative expansion for the
aforementioned processes~\cite{Anastasiou:2015vya,Dulat:2017prg,Mistlberger:2018etf,Cieri:2018oms,Chen:2021isd,Boughezal:2013uia,Chen:2014gva,Boughezal:2015dra,Campbell:2006xx,Campbell:2010cz}. However, one is often
interested in observables where the assumption of a comparatively
large top-quark mass is invalid and the full mass dependence has to be
accounted for. One example is the study of the high-energy tail in the
Higgs boson transverse momentum distribution.

Another avenue towards better theory predictions consists of the
all-order resummation of contributions that are enhanced in kinematic
regions of interest. For Higgs boson production together with at least
one jet, one finds logarithms in $\hat s/\lvert p_\perp \rvert^2$, where $\hat s$
is the square of the partonic centre-of-mass energy and $p_\perp$ a
characteristic transverse momentum scale~\cite{DelDuca:2003ba}. For the case of two or more
jets, the resummation of these high-energy logarithms has been shown
to lead to significant corrections, especially after weak-boson fusion
cuts are applied~\cite{Andersen:2018kjg}. This provides a strong motivation to
extend the study of logarithmic enhancement to the production of a
Higgs boson with a single jet.

The study of logarithmically enhanced high-energy corrections takes different
forms depending on the underlying Born process. For simple, one-scale
processes like Drell-Yan boson production, high-energy corrections arise as
the energy of the hadron collider increases. Such corrections are called
\emph{small-}$x$ \emph{corrections}, since the light-cone momentum fraction
of the incoming partons decreases for increasing hadronic energy. These
corrections are often accounted for using \emph{unintegrated pdfs} and
off-shell scattering matrix elements. The much celebrated BFKL equation can
be used to describe the small-$x$ evolution of the gluon pdfs.

Alternatively, the on-shell scattering involving two or more final state
particles receives logarithmically enhanced perturbative corrections in the
so-called \emph{multi-Regge kinematic limit} of large \emph{partonic} centre
of mass energy $\hat{s}$ and fixed (not growing with $\hat{s}$) and similar transverse
scale $p_\perp$ for the produced particles. As for all other on-shell
scatterings, the perturbative process is calculated with collinear factorised
pdfs. But the BFKL formalism can in this case predict the logarithmic
corrections (in $\hat{s}/\lvert p_\perp \rvert^2$) to the on-shell scattering matrix
elements~\cite{Mueller:1986ey}. The focus of the current study is corrections
of this type.

Inclusive calculations of BFKL resummation for Higgs boson plus jet production
have been performed~\cite{Xiao:2018esv,Celiberto:2020tmb}.  In contrast,
our resummation of high-energy logarithms is based on the \HIGHEJ
(\HEJ) framework~\cite{Andersen:2009nu,Andersen:2009he,Andersen:2011hs,Andersen:2018tnm}. \HEJ provides realistic predictions
through a fully flexible Monte Carlo implementation, supplementing
leading-order perturbation theory with high-energy resummation
retaining exact gauge invariance and momentum conservation. The calculation
presented here is the first time this approach has been used for an inclusive 1-jet
process.  As is
necessary in the high-energy region, the all-order resummation
includes the full effects of finite quark masses. We first review the
formalism and derive the new building blocks required for
leading-logarithmic (LL) resummation for Higgs boson plus jet
production in section~\ref{sec:Hjets_he}. In
section~\ref{sec:comparisons-data}, we compare our predictions to
experimental measurements and propose observables tailored to the
systematic analysis of high-energy corrections. We conclude in
section~\ref{sec:conclusions}.


%% file: Theory.tex
\section{Higgs Boson plus Jets Production in the High-Energy Limit}
\label{sec:Hjets_he}

In the following, we discuss the general properties and structure of
amplitudes in the high-energy limit. We briefly summarise LL
resummation in the \HIGHEJ formalism and derive the new ingredients
for the production of a Higgs boson together with a single jet, and for
processes with two or more jets where the Higgs boson is outside of the jets.

\subsection{Scaling of Amplitudes at High Energies}
\label{sec:scaling}

Generally, we are interested in the behaviour of amplitudes in the
region of \emph{Multi-Regge Kinematics} (MRK). This region is defined
by a large centre-of-mass energy with large invariant masses between
all pairs of outgoing particles with finite transverse momenta. This
is equivalent to a strong ordering in rapidities. Specifically, for a
$2 \rightarrow n$ process, we require
\begin{align}
\label{eq:MRKlimit}
y(p_n) \gg ... \gg y(p_1) \qquad |p_{i\perp}| \sim \text{finite} \; \forall i \in \{1..n\},
\end{align}
where the outgoing particle $i$ has momentum $p_i$, rapidity $y_i \equiv y(p_i)$
and transverse momentum $ |p_{i\perp}|$.

In this region, Regge theory~\cite{Fadin:2006bj} states that the
amplitude should scale as
\begin{equation}
  \label{eq:MRK_scaling}
\mathcal M \sim s_{12}^{\alpha_1(t_1)} ... s_{n \,n+1}^{\alpha_n(t_{n})},
\end{equation}
where the $s_{i\,i+1}$ refers to the invariant mass between particle
$i$ and $i+1$, and $\alpha_i(t_i)$ is the maximum spin of any particle
that can be exchanged in the $t$-channel between particle $i$ and
$i+1$. From this formula, it follows that the leading contribution to
a QCD amplitude is given by the configurations which maximise the
number of gluons exchanged in the $t$-channel. These configurations
characterise the regions of phase space in which the leading
high-energy logarithms arise. We therefore refer to them as
leading-logarithmic (LL) or Fadin-Kuarev-Lipatov (FKL) configurations.

\begin{figure}[htb]
  \centering
  \includegraphics[width=0.45\linewidth]{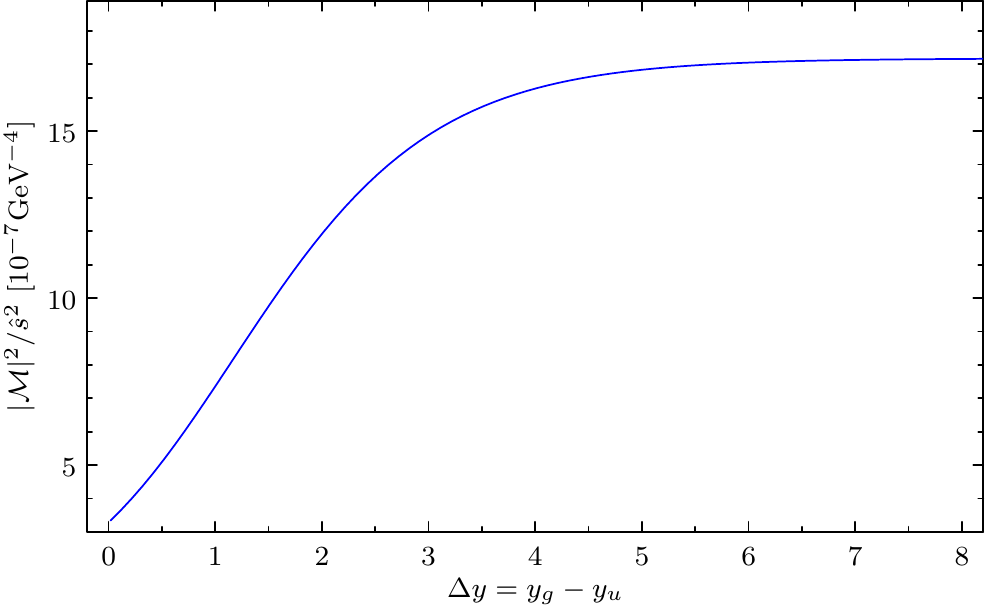}\quad\includegraphics[width=0.45\linewidth]{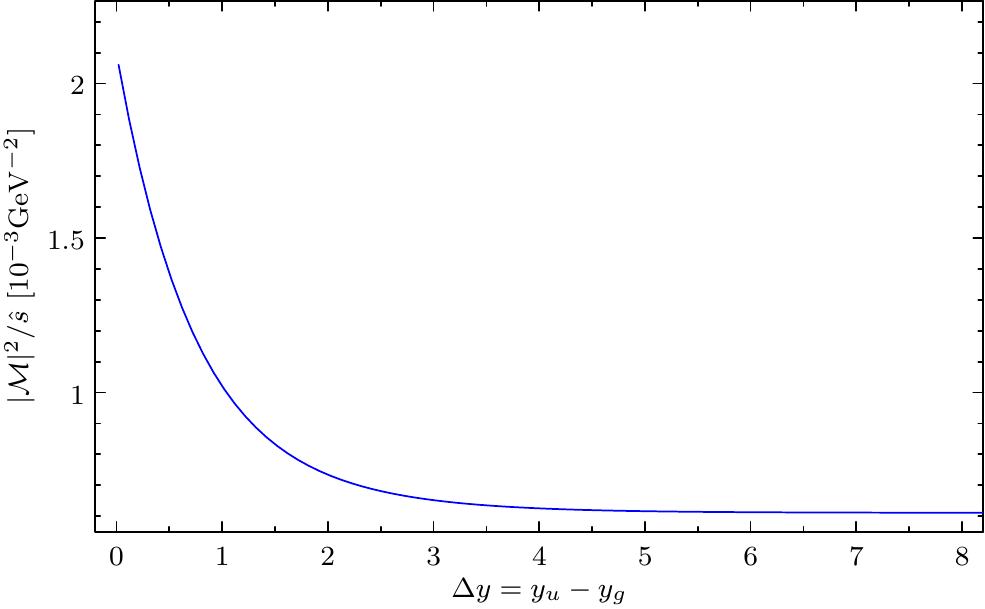}
  \caption{%
    Regge scaling of quark-gluon scattering amplitudes in the MRK limit,
    approached by increasing $\Delta y$.
    Left: Rapidity ordering $qg \to qg$. Right: Rapidity ordering $qg \to gq$. The kinematics are fixed by the azimuthal angle $\phi_1 = \frac{\pi}{7}$ and transverse momentum $p_{1\perp} =40\,$GeV of the outgoing particle in the backward direction.
  }
  \label{fig:MRK_scaling_gq}
\end{figure}

As a simple example, let us consider the amplitude for elastic
scattering of a quark or antiquark ($q$) and a gluon ($g$),
with the incoming quark in the backward direction~\cite{Andersen:2009he}. Ordering the
outgoing particles by ascending rapidity, the two possible
configurations are $qg \to qg$ and $qg \to gq$. For the rapidity
ordering $qg \to qg$ it is possible to exchange a $t$-channel
gluon and we therefore expect the amplitude to scale as
$\mathcal{M}_{qg \to qg} \sim \hat{s}^1$ for $y(p_2) \gg
y(p_1)$. Conversely, the flipped ordering $qg \to gq$ only allows a
$t$-channel (anti-)quark exchange, implying $\mathcal{M}_{qg \to gq} \sim
\hat{s}^{\frac{1}{2}}$. This scaling behaviour is indeed confirmed by
an explicit calculation and illustrated in
figure~\ref{fig:MRK_scaling_gq}, where increasing $\Delta y$ represents
approaching the MRK limit.

\boldmath
\subsection{$H+\ge 2j$ Processes within HEJ}
\unboldmath
\label{sec:h+ge-2j-processes}


The construction of the leading-logarithmic calculation of $pp\to H+\ge2j$ in the \HEJ framework was described in detail
in~\cite{Andersen:2017kfc,Andersen:2018kjg}.  Here we summarise the main points
in order to frame the discussion of the new components calculated in this paper.

Following the arguments in section~\ref{sec:scaling}, the LL
configurations in pure QCD have the form $f_a f_b \to f_a \cdots f_b$,
where $f_a, f_b$ indicate the incoming parton flavours and the
ellipsis denotes an arbitrary number of gluons. As before, the
particles are written in order of increasing rapidity.
The production of an additional Higgs boson proceeds via an effective
coupling to two or more gluons. Since invariant masses are large in
the high-energy region, it is crucial that the exact dependence on the
top-quark mass is included in this effective coupling.
A final-state Higgs boson with momentum $p_H \equiv p_j$ at an intermediate
rapidity $y_j$ such that $y_{j-1} \ll y_j \ll y_{j+1}$ can then exchange
$t$-channel gluons with the outgoing partons $j-1, j+1$. It was shown
in~\cite{Andersen:2017kfc} that the scaling behaviour in
equation~\eqref{eq:MRK_scaling} directly generalises when a Higgs boson is emitted in
the middle of the quarks and gluons.  Therefore, all configurations
$f_a f_b \to f_a \cdots H \cdots f_b$ contribute at LL accuracy.

\begin{figure}[btp]
  \centering
\includegraphics{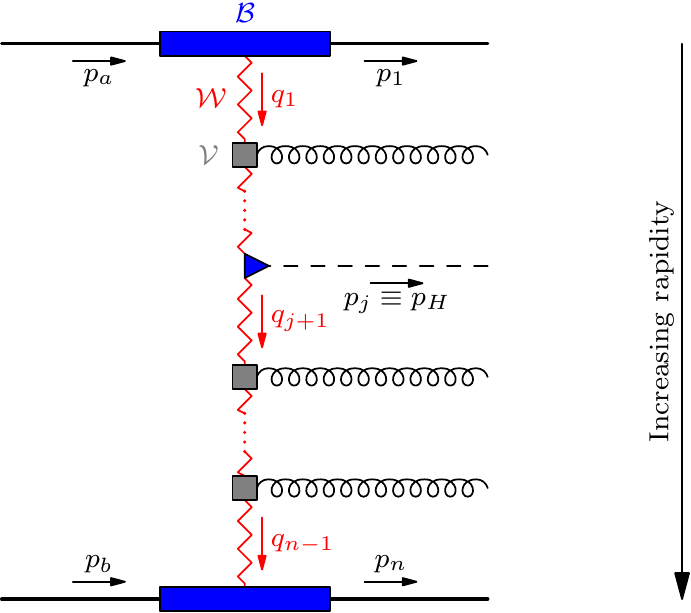}
  \caption{Structure of the matrix element for the process $f_a f_b \to f_a \cdots H \cdots
f_b$.}
  \label{fig:ME_h_jets_central}
\end{figure}

In the MRK limit the amplitudes are found to factorise into a neat product of
simple functions. In
the \HIGHEJ formalism we obtain the form
\begin{equation}
  \label{eq:ME_fact}
  \begin{split}
    \overline{\left|\mathcal{M}_{\HEJ}^{f_a f_b \to f_a \cdots H
          \cdots  f_b}\right|}^2 ={}&\mathcal{B}_{f_a, H, f_b}(p_a, p_b, p_1, p_n, q_j, q_{j+1})\\
    &\cdot \prod_{\substack{i=1\\i \neq j}}^{n-2} \mathcal{V}(p_a,p_b,p_1,p_n, q_i, q_{i+1})\\
    &\cdot \prod_{i=1}^{n-1} \mathcal{W}(q_i, y_i, y_{i+1}),
  \end{split}
\end{equation}
for the modulus square of the matrix element, summed and averaged over helicities
and colours.  In this expression, $p_a$ $(p_b)$ is the
incoming momentum in the backward (forward) direction and
$p_1,\dots,p_n$ are the outgoing momenta ordered in increasing rapidity. The
$t$-channel momenta are given by
\begin{equation}
  \label{eq:q_i}
  q_1 =  p_a - p_1,\qquad q_i = q_{i-1} - p_i \; \text{ for } i > 1.
\end{equation}
The structure is illustrated in figure~\ref{fig:ME_h_jets_central}.
At Born level, the right-hand side of equation~(\ref{eq:ME_fact})
reduces to the function $\mathcal{B}_{f_a, H, f_b}$, described below. $\mathcal{V}$
comprises the real corrections due to the production of $n-3$ gluons
in addition to $f_a, f_b$, and the Higgs boson.
It is given by the contraction
of two Lipatov vertices~\cite{Andersen:2017kfc}:
\begin{align}
  \label{eq:V}
  \mathcal{V}(p_a,p_b,p_1,p_n, q_i, q_{i+1}) ={}& -\frac{C_A}{t_i t_{i+1}} V_\mu(p_a,p_b,p_1,p_n, q_i, q_{i+1}) V^\mu(p_a,p_b,p_1,p_n, q_i, q_{i+1}),\\
  \label{eq:V_Lipatov}
  V^\mu(p_a,p_b,p_1,p_n, q_i, q_{i+1})={}& -(q_i+q_{i+1})^\mu \nonumber\\
  &+ \frac{p_a^\mu}{2} \left( \frac{q_i^2}{p_{i+1}\cdot p_a} +
  \frac{p_{i+1}\cdot p_b}{p_a\cdot p_b} + \frac{p_{i+1}\cdot p_n}{p_a\cdot p_n}\right) +
p_a \leftrightarrow p_1 \nonumber\\
  &- \frac{p_b^\mu}{2} \left( \frac{q_{i+1}^2}{p_{i+1} \cdot p_b} + \frac{p_{i+1}\cdot
      p_a}{p_b\cdot p_a} + \frac{p_{i+1}\cdot p_1}{p_b\cdot p_1} \right) - p_b
  \leftrightarrow p_n,
\end{align}
where $t_i = q_i^2$ are the squares of the $t$-channel momenta.
$\mathcal{W}$ accounts
for the all-order finite contribution coming from the sum of the virtual corrections and
unresolved real corrections.
It is process-independent and described in detail
in~\cite{Andersen:2017kfc}.

The process-dependent Born-level factor is given by
\begin{equation}
  \label{eq:Born_h_jets_central}
 \mathcal{B}_{f_a, H, f_b} =  \frac {(4\pi\alpha_s)^{n-1}} {4(N_c^2-1)}
    \frac {K_{f_a}(p_1^-, p_a^-)} {t_1}\
 \frac{K_{f_b}(p_n^+, p_b^+)}{t_{n-1}} \frac{\left\|S_{f_a
          f_b\to f_a H f_b}\right\|^2}{t_j t_{j+1}}.
\end{equation}
Here, $\alpha_s$ is the strong coupling and $N_c = 3$ is the number of
colours. The difference between incoming gluons and (anti-)quarks is
completely absorbed into the \emph{colour acceleration multipliers}
$K_f$ with
\begin{align}
  \label{eq:K_g}
K_g(x, y) ={}& \frac{1}{2}\left(\frac{x}{y} + \frac{y}{x}\right)\left(C_A -
  \frac{1}{C_A}\right)+\frac{1}{C_A} &\text{for gluons,}\\
  \label{eq:K_q}
  K_q(x, y) ={}& C_F &\text{for quarks and antiquarks.}
\end{align}
$C_F = \frac{N_C^2-1}{2N_C}$ and $C_A = N_C$ are the usual Casimir
invariants. $S_{f_a f_b\to f_a H f_b}$ is a contraction of currents
with the Higgs boson production vertex. The double vertical bars
indicate the sum over helicities of the corresponding amplitudes:
\begin{equation}
  \label{eq:S_fHf}
  \left\|S_{f_a f_b\to f_a H f_b}\right\|^2 = \sum_{\substack{\lambda_a =
  +,-\\\lambda_b = +,-}} \left|j^{\lambda_a}_\mu(p_1, p_a) V_H^{\mu\nu}(q_j,q_{j+1})
j_\nu^{\lambda_b}(p_n, p_b)\right|^2.
\end{equation}
$V_H$ is the well-known one-loop effective coupling between the Higgs boson and
two gluons in the normalisation of~\cite{Andersen:2018kjg}, including the full
quark-mass dependence. The inclusion of this piece in
equation~\eqref{eq:ME_fact} then gives the correct finite quark-mass
contributions at LL for \emph{any} number of final state partons/jets.  Finally,
the current $j$ is given by
\begin{equation}
  \label{eq:current}
  j^{\lambda}_\mu(p, q) = \bar{u}^\lambda(p)\gamma_\mu u^\lambda(q).
\end{equation}

In addition to the LL resummation discussed so far, gauge-invariant
subsets of next-to-leading logarithmic (NLL) corrections originating
from non-FKL configurations have also been included in \HEJ. One
source of NLL corrections are the configurations $qf_b \to Hq \cdots
f_b$ and $f_aq \to f_a \cdots qH $, which only permit $n-2$
$t$-channel gluon exchanges instead of the $n-1$ exchanges found in LL
configurations. In these cases, we adapt the matrix
element formula for the corresponding LL configurations to a flipped
rapidity order of outgoing (anti-)quark and Higgs boson. If the Higgs boson
is emitted first in rapidity order, we use equation~(\ref{eq:ME_fact})
with $p_2 \equiv p_H$ and exclude the virtual correction factor
$\mathcal{W}$ for $i=1$. In the other case of the Higgs boson being
emitted last, we set $p_{n-1}=p_H$ and skip $\mathcal{W}$ for $i=n-1$.

A second class of non-FKL configurations arises for three or more
produced jets, when the most backward or forward outgoing particle is
a gluon, but the corresponding incoming parton is a quark or
antiquark. These ``unordered gluon'' configurations, $q f_b \to
gq\cdots H \cdots f_b$ and $f_a q \to f_a \cdots H \cdots q g$, allow
one $t$-channel gluon exchange less than the corresponding FKL
configurations in which the unordered gluon is swapped with the
neighbouring (anti-)quark. Hence, they contribute at NLL
accuracy. Without loss of generality, we consider the case where the
unordered gluon is the most backward emitted particle. We denote its
momentum by $p_g$ and the following momenta by $p_1,\dots,p_n$. The
modulus square of the matrix element then has the same structure as in
equation~(\ref{eq:ME_fact}). In fact, the only changes are that the
first $t$-channel momentum is now $q_1 = p_a - p_1 - p_g$ and that a
different Born-level function $\mathcal{B}_{gq,H,f_b}$ depending also
on $p_g$ appears. For a derivation and explicit expressions,
see~\cite{Andersen:2017kfc}.

\boldmath
\subsection{Scaling of  $H+\ge 1j$ Amplitudes}
\label{sec:scaling_h1j}
\unboldmath

To extend the formalism to the production of a Higgs boson with a
single jet we first need to identify the LL configurations, following
the discussion in section~\ref{sec:scaling}, and then derive the
corresponding matrix elements.

So far, we have only considered LL configurations in which both the most backward and the most forward outgoing particle is a parton. However, in the process $g q \rightarrow H q$, the amplitude should scale as $\mathcal M \sim s_{Hq}$, as there is a gluon exchange (thus a spin-1 particle) in the $t$-channel. Similarly, the process $g g \rightarrow H g$ corresponds to $\mathcal M \sim s_{Hg}$. If we look at Higgs boson plus dijet production, the same argument allows us to establish that $g q \rightarrow H g q$ scales as $\mathcal M \sim s_{Hg} s_{gq}$. All these configurations therefore contribute at LL accuracy. This is no longer the case if, for example, outgoing parton flavours are rearranged: $g q \rightarrow H q g$ scales as $\mathcal M \sim s_{Hq} s_{qg}^{1/2}$.

\begin{figure*}[h!]
        \centering
        \begin{subfigure}[b]{0.475\textwidth}
            \centering
            \includegraphics[width=\textwidth]{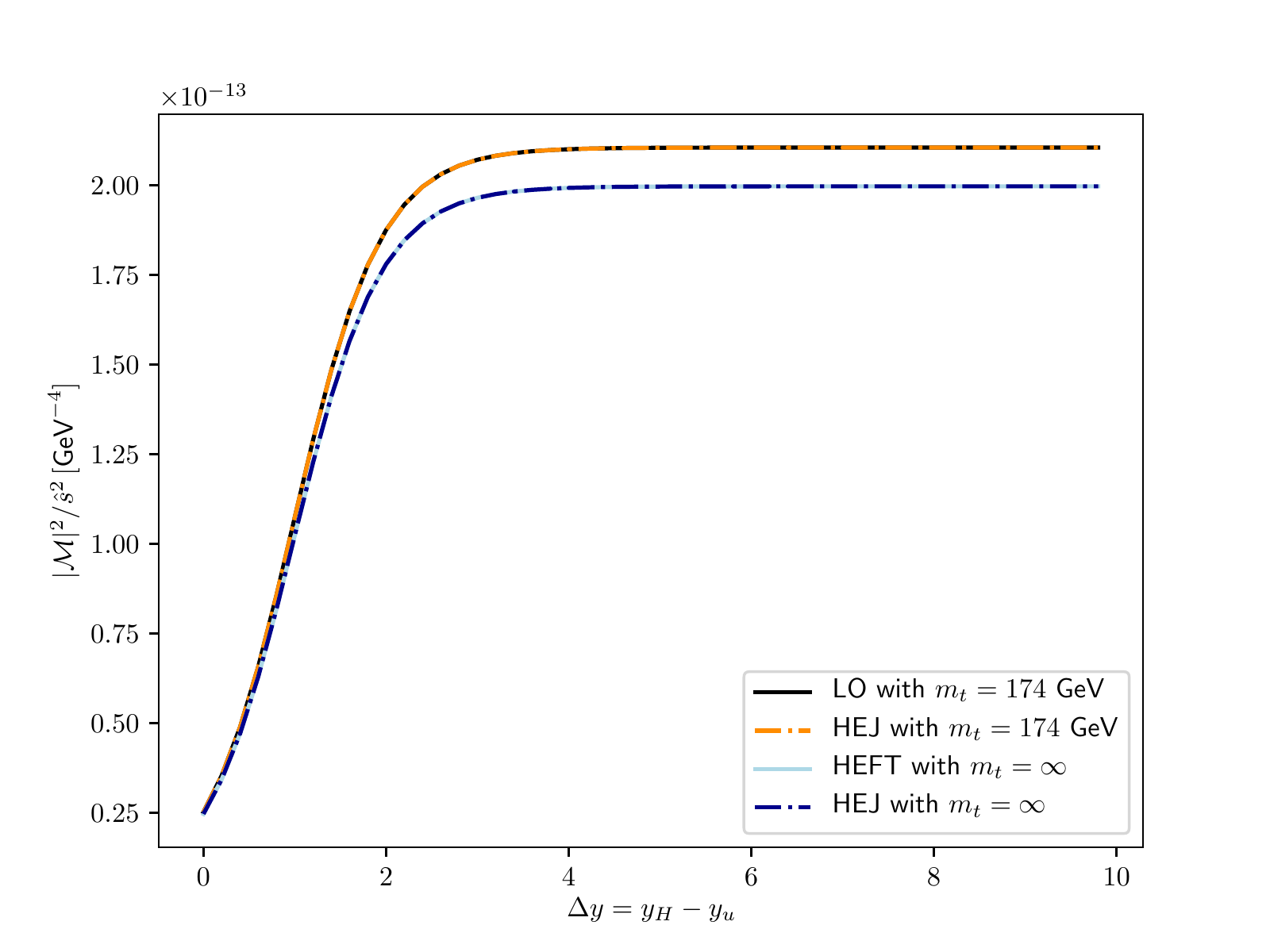}
            \caption[Network2]%
            {{\small $|\mathcal M|^2/\hat s^2$ for the process $g u \rightarrow H u$}}
            \label{fig:Hq}
        \end{subfigure}
        \hfill
        \begin{subfigure}[b]{0.475\textwidth}
            \centering
            \includegraphics[width=\textwidth]{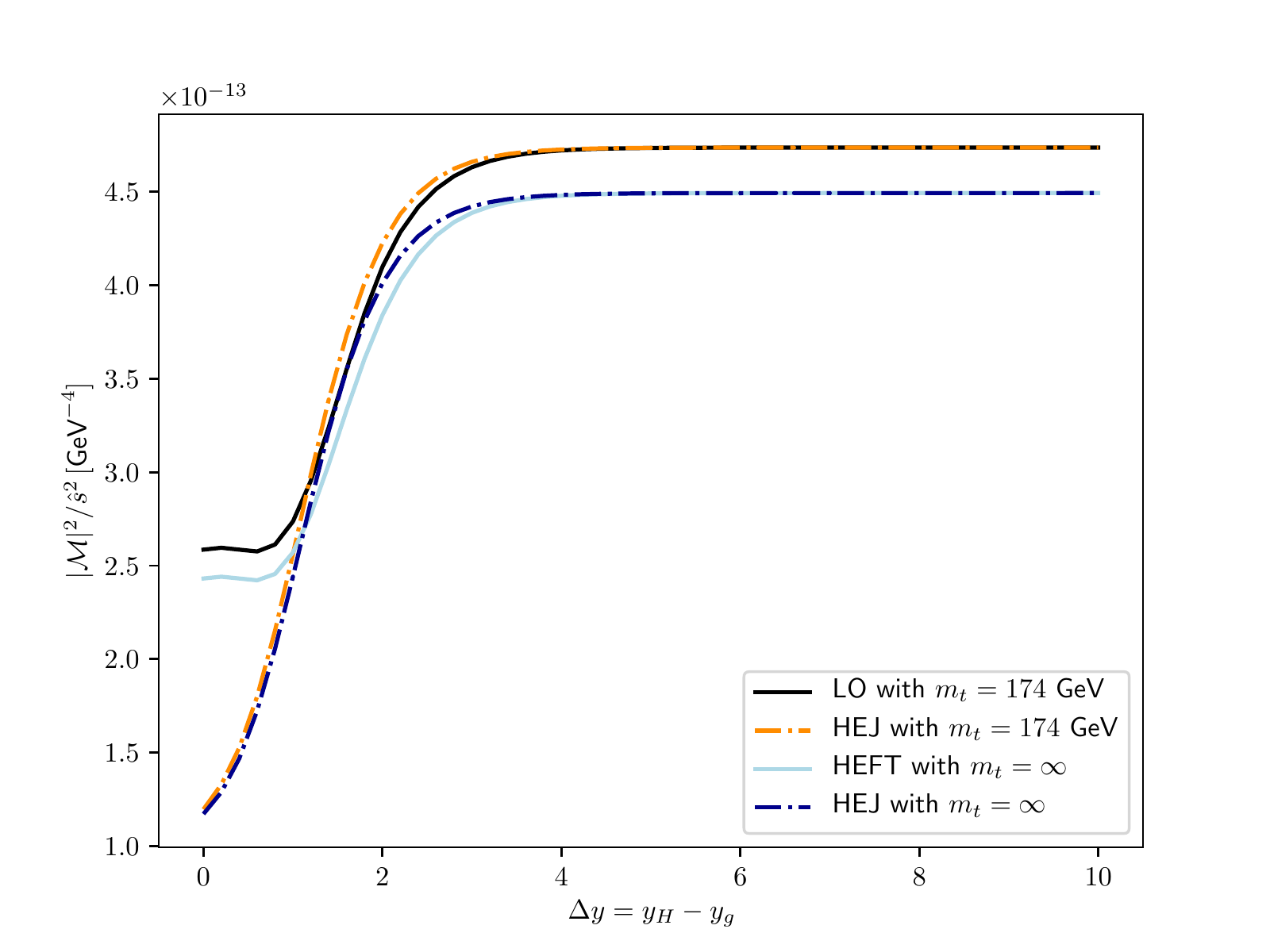}
            \caption[]%
            {{\small $|\mathcal M|^2/\hat s^2$ for the process $g g \rightarrow H g$}}
            \label{fig:Hg}
        \end{subfigure}
        \vskip\baselineskip
        \begin{subfigure}[b]{0.475\textwidth}
            \centering
            \includegraphics[width=\textwidth]{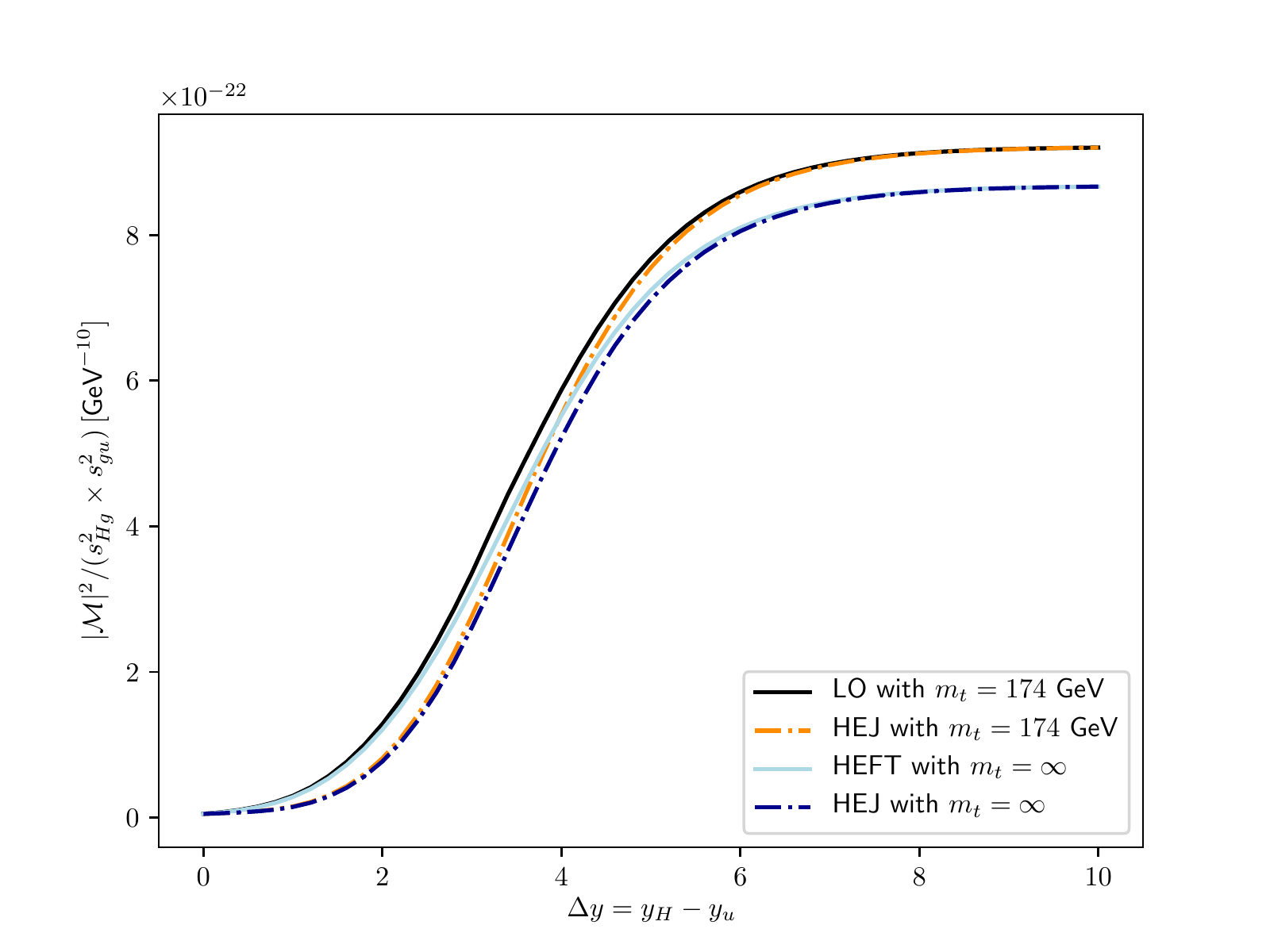}
            \caption[]%
            {{\small $|\mathcal M|^2/(s_{Hg}^2 \times s_{gu}^2))$ for the process $g u \rightarrow H g u$}}
            \label{fig:Hgq}
        \end{subfigure}
        \hfill
        \begin{subfigure}[b]{0.475\textwidth}
            \centering
            \includegraphics[width=\textwidth]{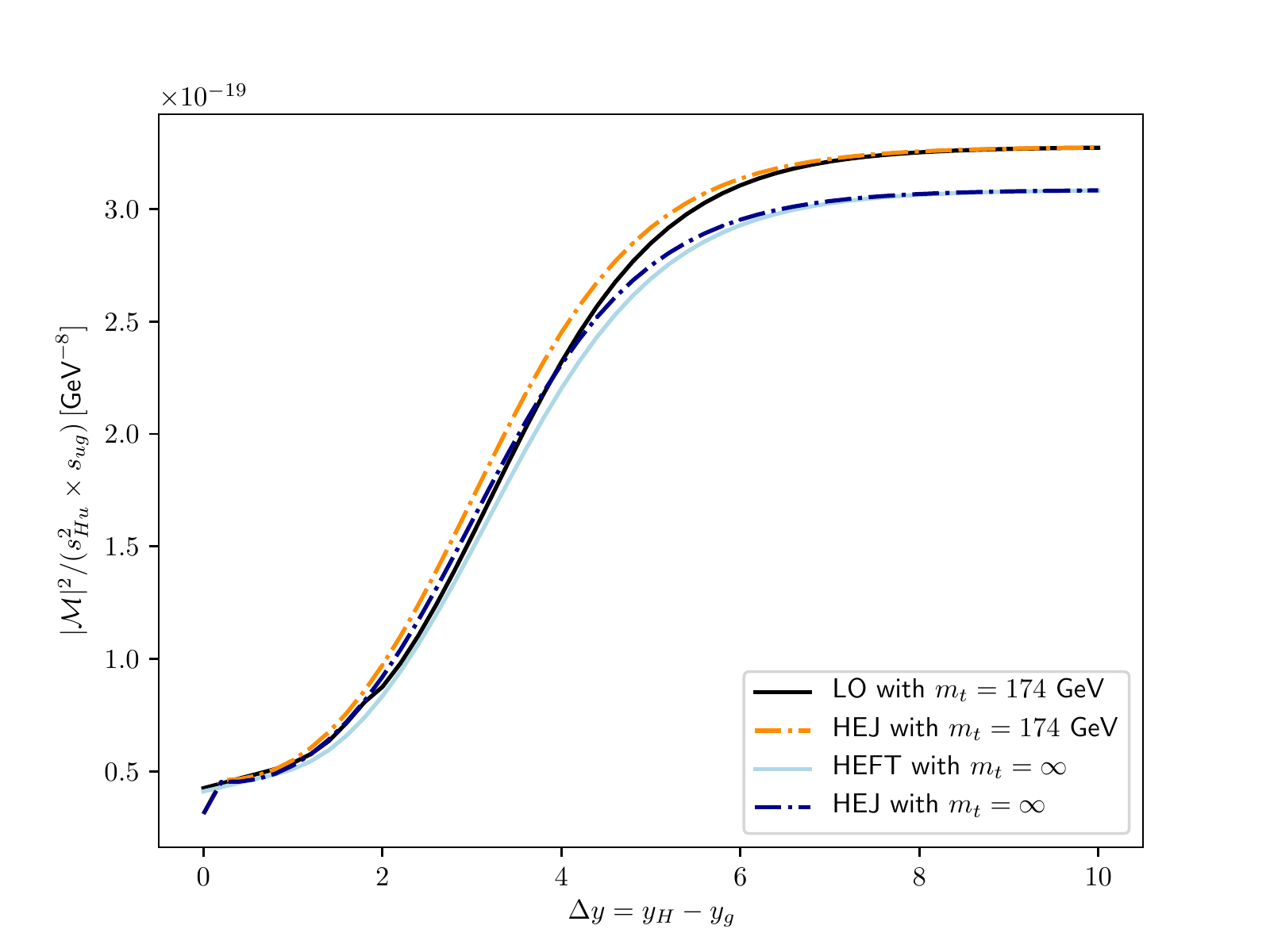}
            \caption[]%
            {{\small $|\mathcal M|^2/(s_{Hu}^2 \times s_{ug}))$ for the process $g u \rightarrow H u g$}}
            \label{fig:Hqg}
        \end{subfigure}
        \caption[ ]
        {\small Verifying Regge scaling of the squared matrix elements
          (equation~\eqref{eq:MRK_scaling}) for 4 different processes.
          Increasing values of $\Delta y$ on the $x$-axis approach
          the MRK limit (equation~\eqref{eq:MRKlimit}).}
        \label{fig:scaling_explorers}
    \end{figure*}

Note that these scalings are valid whether we consider the full LO amplitude (with Higgs to gluons couplings \emph{via} quark loops) or the Higgs Effective Field Theory (HEFT) one with an infinite top mass $m_t$, as shown in figure~\ref{fig:scaling_explorers}. To produce these plots, the amplitude is extracted from Madgraph5\_aMC@NLO~\cite{Alwall:2014hca} and is calculated in a one-dimensional phase-space as a function of the rapidity separation between all pairs of particles. It was checked for consistency that setting the top mass to infinity in the LO amplitude yields the HEFT result. We compare to the LO truncation of the all-order \HEJ amplitudes, anticipating their derivation from the high-energy limit in section~\ref{sec:lipatov-vertex-an},

The momentum configurations chosen are summarised in
table~\ref{tab:explorerPS}.  We stress though that the behaviour shown is not
dependent on specific values of azimuthal angle or transverse momentum, but only on the rapidity assignment of the particles.

\begin{table}[H]
\begin{center}
\begin{tabular}{ |c||l| }
\hline
Process & Momenta configuration \\
\hline
$gq \rightarrow Hq$ &
\(
\begin{cases}
      y_q = -\Delta \text{ and } y_H = \Delta &\\
      \phi_q = \frac{\pi}{4} & \\
     p_{q\perp} = 40 \, \text{GeV}  &
    \end{cases}
\)  \\
$gg \rightarrow Hg$ &
\(
\begin{cases}
      y_g = -\Delta \text{ and } y_H = \Delta &\\
      \phi_g = \frac{\pi}{4} & \\
     p_{g\perp} = 40 \, \text{GeV}  &
    \end{cases}
\)     \\
$gq \rightarrow Hgq$ &
\(
\begin{cases}
      y_q = -\Delta, y_g=0 \text{ and } y_H = \Delta &\\
      \phi_g = \frac{\pi}{2} \text{ and } \phi_q =-\frac{\pi}{3} & \\
     p_{g\perp} = k_{q\perp} = 40 \, \text{GeV}  &
    \end{cases}
\)  \\
$gq \rightarrow Hqg$ &
\(
\begin{cases}
      y_g = -\Delta, y_q=0 \text{ and } y_H = \Delta &\\
      \phi_g = -\frac{\pi}{2} \text{ and } \phi_q = \frac{\pi}{3} & \\
     p_{g\perp} = k_{q\perp} = 40 \, \text{GeV}  &
    \end{cases}
\)  \\
 \hline
\end{tabular}
\end{center}
\caption{The momentum configurations used in figure~\ref{fig:scaling_explorers}.}
\label{tab:explorerPS}
\end{table}

\boldmath
\subsection{New Components for $H+\ge 1j$ and an Outer Higgs Boson}
\label{sec:lipatov-vertex-an}
\unboldmath

In section~\ref{sec:scaling} we discussed the factorisation of LL
amplitudes for $f_a f_b \to f_a \cdots H \cdots f_b$ into a Born-level
function $\mathcal{B}$, a product over real-emission vertices
$\mathcal{V}$, and a product of virtual corrections $\mathcal{W}$. The
same type of factorisation holds for LL configurations with the Higgs
boson as the most forward or backward outgoing particle. In fact, the
virtual corrections are the same as in equation~(\ref{eq:ME_fact}). To
derive the remaining factors, we first analyse the Born-level process
$gf_b \to Hf_b$ and then consider real corrections.

\subsubsection{Higgs Current}
\label{sec:higgs-current}

\begin{figure}[htb]
  \centering
  \includegraphics{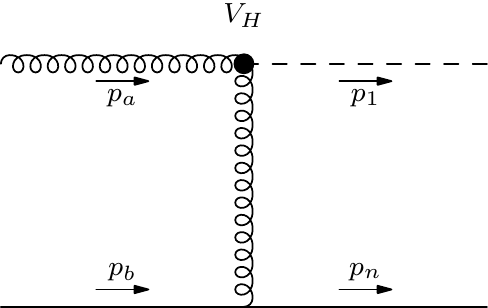}
  \caption{The tree-level diagram for the process $gq\to Hq$. The
straight solid line denotes an arbitrary light quark or antiquark.}
  \label{fig:gq_to_Hq}
\end{figure}

The Born-level function $\mathcal{B}_{H,f_b}$ for the process $gf_b
\to H \cdots f_b$ is obtained by deriving a $t$-channel factorised form
analogous to equation~(\ref{eq:Born_h_jets_central}) from the modulus square
of the Born-level amplitude in the MRK limit. For $gq \to H q$, the
tree-level amplitude is determined by a single diagram, depicted in
figure~\ref{fig:gq_to_Hq}.

Without requiring any approximations we
obtain the factorised expression
\begin{align}
  \label{eq:B_Hq}
  \mathcal{B}_{H,f_b} ={}& \frac {(4\pi\alpha_s)^{n-1}} {4(N_c^2-1)}
                           \frac {1} {t_1}
                           \frac{K_{f_b}(p_n^+, p_b^+)}{t_{n-1}}
                           \left\|S_{g f_b \to H f_b}\right\|^2,\\
  \label{eq:S_Hf}
  \left\|S_{g f_b \to  H f_b}\right\|^2 ={}&  \sum_{
  \substack{
  \lambda_{a} = +,-\\
  \lambda_{b} = +,-
  }}
 \left|\epsilon_\mu^{\lambda_a}(p_a)\ V_H^{\mu\nu}(p_a, p_a-p_1)\ j_\nu^{\lambda_b}(p_n, p_b)\right|^2,
\end{align}
for $f_b = q$, where $\epsilon^{\lambda_a}(p_a)$ is the polarisation
vector of the incoming gluon. This is plotted along with the exact LO
results from Madgraph5\_aMC@NLO~\cite{Alwall:2014hca} in
figure~\ref{fig:scaling_explorers}(a), showing exact agreement for
both finite top quark mass and in the infinite $m_t$ limit.  In the
MRK limit, this formula also holds for $f_b = g$, which is shown in
figure~\ref{fig:scaling_explorers}(b).  In this case there is some
approximation away from the limit, but very quickly the LO and HEJ
lines converge as $\Delta y$ increases.

\subsubsection{Lipatov Vertex for Additional Gluons}
\label{sec:lipat-vert-addit}

In section~\ref{sec:h+ge-2j-processes}, we described the simple factorised
structure of amplitudes within (N)MRK limits.  Not only are the different
components independent of momenta in different parts of the chain, they are
independent of the particle content of the rest of the chain.  This should mean
that the Lipatov vertex derived in pure QCD processes for additional gluons
still applies.  However, the Lorentz and colour structure of the
``Higgs current'' $j_H^\nu = \epsilon_\mu V_H^{\mu\nu}$ differ
compared to pure QCD processes so it is important to check that this
is indeed the case.

\begin{figure}[H]
  \centering
  \begin{tabular}{cccc}
    \includegraphics[width=0.2\textwidth]{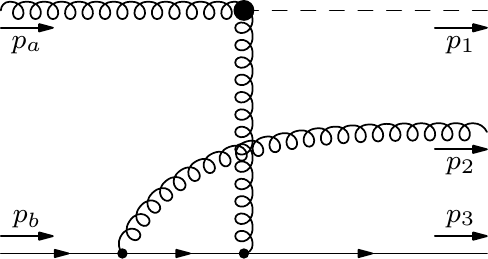}
    &\includegraphics[width=0.2\textwidth]{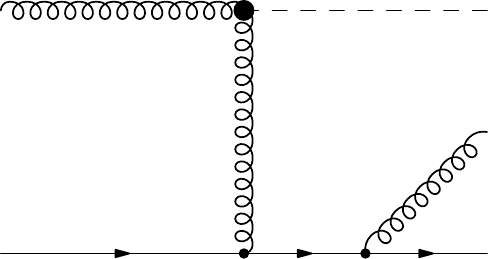}
    &\includegraphics[width=0.2\textwidth]{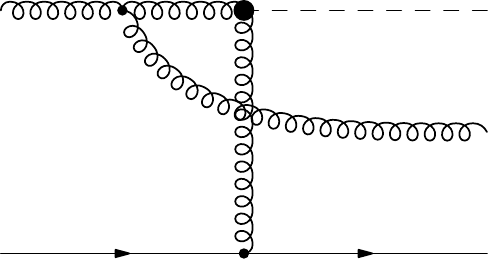}
    &\includegraphics[width=0.2\textwidth]{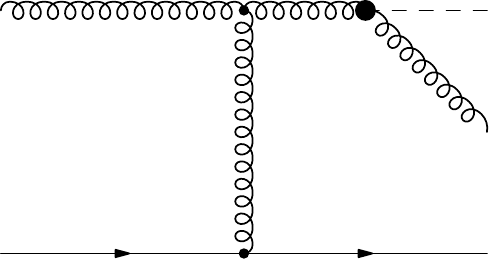}\\
    a) & b) & c) & d)
  \end{tabular}\\[1em]

  \begin{tabular}{cccc}
    \includegraphics[width=0.2\textwidth]{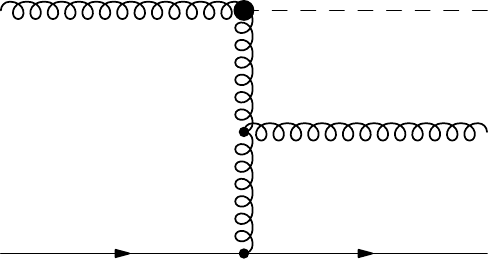}
    &\includegraphics[width=0.2\textwidth]{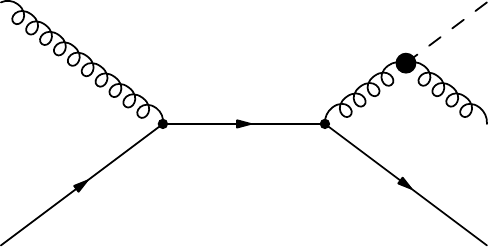}
    &\includegraphics[width=0.2\textwidth]{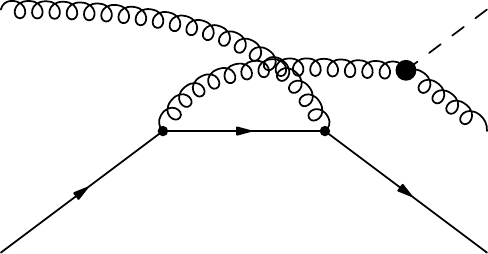}
    &\includegraphics[width=0.2\textwidth]{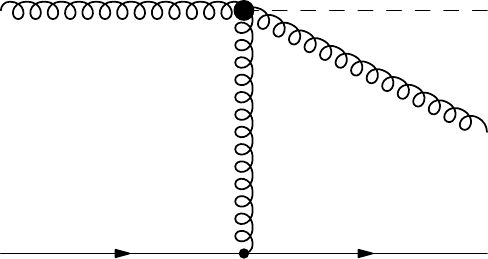}\\
    e) & f) & g) &h)
  \end{tabular}
  \caption{The eight LO diagrams which contribute to the process in
    equation~(\ref{eq:Hgq}).}
  \label{fig:LOhjj}
\end{figure}

We will consider the process
\begin{align}
  \label{eq:Hgq}
  g(p_a) q(p_b) \rightarrow H(p_1) g(p_2) q(p_3),
\end{align}
in the MRK limit $y_1\ll y_2 \ll y_3$. There are eight LO diagrams, as shown in
fig.~\ref{fig:LOhjj}.  Compact expressions for tree-level Higgs-plus-4 parton
colour-ordered amplitudes appear in~\cite{Dawson:1991au,Kauffman:1996ix}.  Setting $q_1 = p_a - p_1$ and $q_2 = p_3 - p_b$, the \HEJ amplitude is given by
\begin{align}
  \label{eq:MHEJlip}
  \begin{split}
    \mathcal{M}_{\HEJ} = i g_s^2 f^{2ea}T^e_{3b}\ \frac{\bar{u}(p_3) \gamma^\nu
      u(p_b)}{q_1^2 q_2^2}\ V^\alpha(p_a, p_b, p_a, p_3, q_1, q_2) V_H^{\mu\nu}(p_a,
    q_1)\ \varepsilon_\mu(p_a) \varepsilon_{\alpha}(p_2)^*.
  \end{split}
\end{align}
As the outer particle is no longer colour-charged, the third argument of the Lipatov vertex defined in equation~\eqref{eq:V_Lipatov} is now $p_a$ instead of $p_1$. The
colour factor of the \HEJ amplitude may be rewritten
\begin{align}
  \label{eq:colord}
  if^{2ea}T^e_{3b} = (T^a T^2)_{3b} - (T^2 T^a)_{3b}.
\end{align}
We can then directly compare equation~\eqref{eq:MHEJlip} with the MRK limit of eqs.~(26)
and (27) in Ref.~\cite{Kauffman:1996ix}, and we find agreement at LL up to an
unphysical phase arising from our spinor conventions.  Specifically, the LL term
in the MRK and infinite top-quark mass limit of equation~\eqref{eq:MHEJlip} is given by
\begin{align}
  \label{eq:LipatovLimit}
  g_s^2\left( \frac{\alpha_s}{3\pi v} \right)
  if^{2ea}T^e_{3b}
  \frac{\langle 3a \rangle [a b]}{|p_{2\perp}| |p_{3\perp}|} \rightarrow
  g_s^2 \left( \frac{\alpha_s}{3\pi v} \right)
  if^{2ea}T^e_{3b}
  \frac{\hat s}{|p_{2\perp}| |p_{3\perp}|},
\end{align}
where the angle and square brackets are Lorentz-invariant kinematic factors defined by $\langle ij \rangle = \bar{u}(p_i)P_Ru(p_j)$ and $[ij] = \bar{u}(p_i)P_Lu(p_j)$.

\subsubsection{Matrix element including additional gluons}
\label{sec:matr-elem-incl}
We can now use these results to form the analogue of equation~\eqref{eq:ME_fact} for
the process $gf_b\to H \cdots f_b$
\begin{align}
  \label{eq:ME_fact_outer}
  \begin{split}
    \overline{\left|\mathcal{M}_{\HEJ}^{g f_b \to H \cdots
           f_b}\right|}^2 ={}&\mathcal{B}_{ H, f_b}(p_a, p_b, p_1, p_n)\\
    &\cdot \prod_{i=1}^{n-2} \mathcal{V}(p_a, p_b, p_a, p_n, q_i, q_{i+1})\\
    &\cdot \prod_{i=1}^{n-1} \mathcal{W}(q_i, y_i, y_{i+1}),
  \end{split}
\end{align}
where the only differences to equation~(\ref{eq:ME_fact}) are the
Born-level function $\mathcal{B}_{ H, f_b}(p_a, p_b, p_1, p_n)$ given in equation~\eqref{eq:B_Hq} and the third argument of the
real-correction function $\mathcal{V}$.  We illustrate that this gives the
correct behaviour in the MRK limit in figure~\ref{fig:scaling_explorers}(c) for
the processes $gu\to Hgu$, and in figure~\ref{fig:scaling_explorers}(d) we show
that we obtain the correct limiting behaviour for the NLL configuration $gu\to Hug$.

%% file: ComparisonData.tex
\section{Predictions and Comparison to Data}
\label{sec:comparisons-data}
In this section we compare predictions for Higgs boson production in
association with one or more jets obtained with High Energy Jets to those of
fixed next-to-leading order perturbation theory and to experimental
analyses. The analyses are implemented in Rivet~\cite{Bierlich:2019rhm} and
relate to data collected at the LHC operated at both 13~TeV\cite{CMS:2018ctp,CMS:2022wpo}
and 8~TeV~\cite{ATLAS:2014yga}.

\subsection{Predictions}

In our predictions, Sherpa~\cite{Bothmann:2019yzt} is used to generate
leading-order events through Comix~\cite{Gleisberg_2008} and
Openloops~\cite{Buccioni:2017yxi} for $H+n$ jets, where $n=1,\dots, 5$. We
include the exact dependence on the top-quark mass where available (i.e.~for $n=1,2$) and for higher
multiplicities use the simpler results valid for an infinite top mass.
High-energy resummation is then applied using the method of \HEJ2, described
in detail in~\cite{Andersen:2018tnm}.  This takes the fixed-order events as
input and then adds all-order corrections (real and virtual) corresponding to
each Born phase space point. The resulting resummation events are reweighted by
\begin{align}
  \label{eq:reweight_ME}
  \frac{|{\cal M}_{\text{\HEJ}}(m_t,m_b)|^2}{|{\cal M}_{\text{\HEJ, LO}}(m_t,0)|^2}& &\leq 2 \text{ jets},\\
  \frac{|{\cal M}_{\text{\HEJ}}(m_t,m_b)|^2}{|{\cal M}_{\text{\HEJ, LO}}(\infty, 0)|^2}& &> 2 \text{ jets}.
\end{align}
${\cal M}_{\text{\HEJ}}(m_t,m_b)$ is the \HEJ all-order matrix element
discussed in section~\ref{sec:Hjets_he}, where we have indicated the
dependence on the top-quark mass $m_t$ and the bottom-quark mass
$m_b$. ${\cal M}_{\text{\HEJ, LO}}(m_t,m_b)$ denotes the leading-order
truncation of the \HEJ matrix element. The $p_T$-sampling for the leading-order
events used for the matching extends slightly beyond the cuts used in the
analysis, as required by the mapping between the high-multiplicity $m$-body
resummation phase space point and the $n$-parton ($n<m$) phase space point of
the matching. One way to look at this is that the radiation produced by the
resummation on top of the fixed-order input modifies the momenta in the
input, and the over-sampling is needed in order for the full
analysis-phase space of the resummation events to be covered.

We also use Sherpa and Openloops to provide NLO 1-jet and 2-jet predictions
in the infinite top-quark mass limit without resummation, for comparisons
with \HEJ and the experimental data.
The cross sections presented from \HEJ are further matched to NLO by
multiplying the predictions for the inclusive 1-jet (or 2-jet) distributions
by the ratio of the inclusive 1-jet (resp. 2-jet) cross-section at NLO
divided by the inclusive 1-jet (resp. 2-jet) cross-section of HEJ expanded to
NLO. This changes the normalisation of distributions, and reduces the scale
variation.
\begin{align}
  \label{HEJNLONJ}
  \frac{d\sigma_{\text{HEJNLO}n\text{J}}}{d\mathcal{O}} ={}&
  \frac{\sigma_{\text{NLO}n\text{J}}} {\sigma_{\text{\HEJ}n\text{J}}}\frac{d\sigma_{\text{\HEJ}}}{d\mathcal{O}},
\end{align}
where $\sigma_{\text{NLO}n\text{J}}, n=1,2$ denotes the inclusive $n$-jet
cross section at NLO and $\sigma_{\text{\HEJ}n\text{J}}$ the \HEJ
prediction for the inclusive $n$-jet cross section.  Note that the components
of the cross section with exclusive three or more jets as predicted by \HEJ
are technically matched only at Born level, but since they form part of the
inclusive one or two-jet observables, their contribution is scaled by the
relevant ratio in eq.~\eqref{HEJNLONJ}.

We use the NNPDF30@NNLO~\cite{Ball_2015} PDF set provided from the LHAPDF collaboration~\cite{Buckley_2015} for \HEJ and NLO predictions, with the central scale choice $\mu_F =
\mu_R = \max(m_{12}, m_H)$ (where $m_{12}$ is the invariant mass between the two
hardest jets, and set to $m_{12}=0$ for 1-jet events). In order to gauge the
scale dependence of the predictions the scales are varied independently by a
conventional factor of two, excluding combinations where $\mu_F$ and $\mu_R$
differ by a factor of more than two. The coloured regions in the figures below
indicate the theoretical uncertainty envelope formed by these scale variations.

We also investigated an alternative central scale choice $\mu_F = \mu_R = H_T/2$.
The predictions changed only minimally with this scale compared to the custom
scale choice above and so are not presented in this study.

\subsection{Predictions for 13 TeV and Comparison to Data}
\label{sec:13TeV}

In this section we present predictions for a CMS
analysis~\cite{CMS:2018ctp,CMS:2022wpo} at a centre-of-mass energy of
$\sqrt{\hat{s}} = 13$ TeV
and for additional distributions showcasing differences between \HEJ and fixed
order predictions at NLO. The CMS study explored distributions for Higgs boson
production (and decay in the di-photon channel) both inclusively and in
association with one jet.

The baseline cuts related to the photons and the jets are listed in
table~\ref{table:13TeV-baseline} (see refs.~\cite{CMS:2018ctp,CMS:2022wpo} for a full
discussion). The pseudo-rapidity jet cuts are specific to the observables studied and are listed in table~\ref{table:13TeV-baseline-observables}. Jets are reconstructed with the anti-$k_T$~\cite{Cacciari:2008gp}
jet algorithm with $R=0.4$.

\begin{table}[hbt]
\begin{center}
\begin{tabular}{ |l|l| }
 \hline
Description & Baseline cuts  \\
 \hline
 Leading photon transverse momentum & $p_T(\gamma_1) > 30 \text{ GeV}$  \\
 Subleading photon transverse momentum & $p_T(\gamma_2) > 18 \text{ GeV}$  \\
 Diphoton invariant mass & $m_{\gamma\gamma} > 90 \text{ GeV} $ \\
 Pseudo-rapidity of the photons & $|\eta_{\gamma} |<2.5$ \\ & excluding $1.4442< |\eta_\gamma| < 1.566$ \\
 Ratio of harder photon $p_T$ to diphoton invariant mass & $p_T(\gamma_1)/m_{\gamma \gamma} > \frac{1}{3} $  \\
 Ratio of softer photon $p_T$ to diphoton invariant mass & $p_T(\gamma_2)/m_{\gamma \gamma} > \frac{1}{4} $  \\
 Photon isolation cut & $\text{Iso}^\gamma_\text{gen} < 10 \text{ GeV}$ \\
 Jet transverse momentum & $p_T(j) > 30 \text{ GeV}$  \\
 \hline
\end{tabular}
\caption{Baseline photon and jet cuts of the 13 TeV analysis, following the CMS analysis of~\cite{CMS:2018ctp,CMS:2022wpo}. $\text{Iso}^\gamma_\text{gen}$ denotes the sum of transverse energies of stable particles in a cone of radius $\Delta R$ = 0.3 around each photon.}
\label{table:13TeV-baseline}
\end{center}
\end{table}

\begin{table}[hbt]
\begin{center}
\begin{tabular}{ |l|l| }
 \hline
 Observable & Pseudo-rapidity jet cut \\
 \hline
 Number of jets $N_{\text{jets}}$, figure~\ref{fig:13TeV_CS_Njets} & $|\eta_j| <2.5$ (all jets)  \\
 $|p_{T}^{j_1}|$, figure~\ref{fig:13TeV_CS_pTj1} & $|\eta_{j_1}| <2.5$ (hardest jet) and $|\eta_j| <4.7$ (other jets)  \\
 $\min m_{ff}$, figure~\ref{fig:13TeV_CS_minDyany} & $|\eta_{j_1}| <2.5$ (hardest jet) and $|\eta_j| <4.7$ (other jets)  \\
 $\max m_{ff}$, figure~\ref{fig:13TeV_CS_maxinvmassany} & $|\eta_{j_1}| <2.5$ (hardest jet) and $|\eta_j| <4.7$ (other jets)  \\
\hline
\end{tabular}
\caption{Pseudo-rapidity jet cuts used for the 13 TeV analysis observables presented in this section, following the CMS analysis of~\cite{CMS:2018ctp,CMS:2022wpo}.}
\label{table:13TeV-baseline-observables}
\end{center}
\end{table}

The \HEJ and NLO QCD predictions only describe $pp\to H+n$-jet processes via gluon fusion (GF) where the
jets consist of light quarks and gluons. The data includes a non-GF contribution from
electroweak VBF, $VH$ and $t\bar tH$ processes, labelled together as $HX$ in the
experimental papers.  We have extracted the value of this component from the
experimental papers for the rest of this section, and added it to both the \HEJ
and NLO QCD predictions, where possible.  This is indicated with ``+HX'' in the legend.

\begin{figure}[H]
        \begin{subfigure}[b]{0.475\textwidth}
            \centering
            \includegraphics[width=\textwidth]{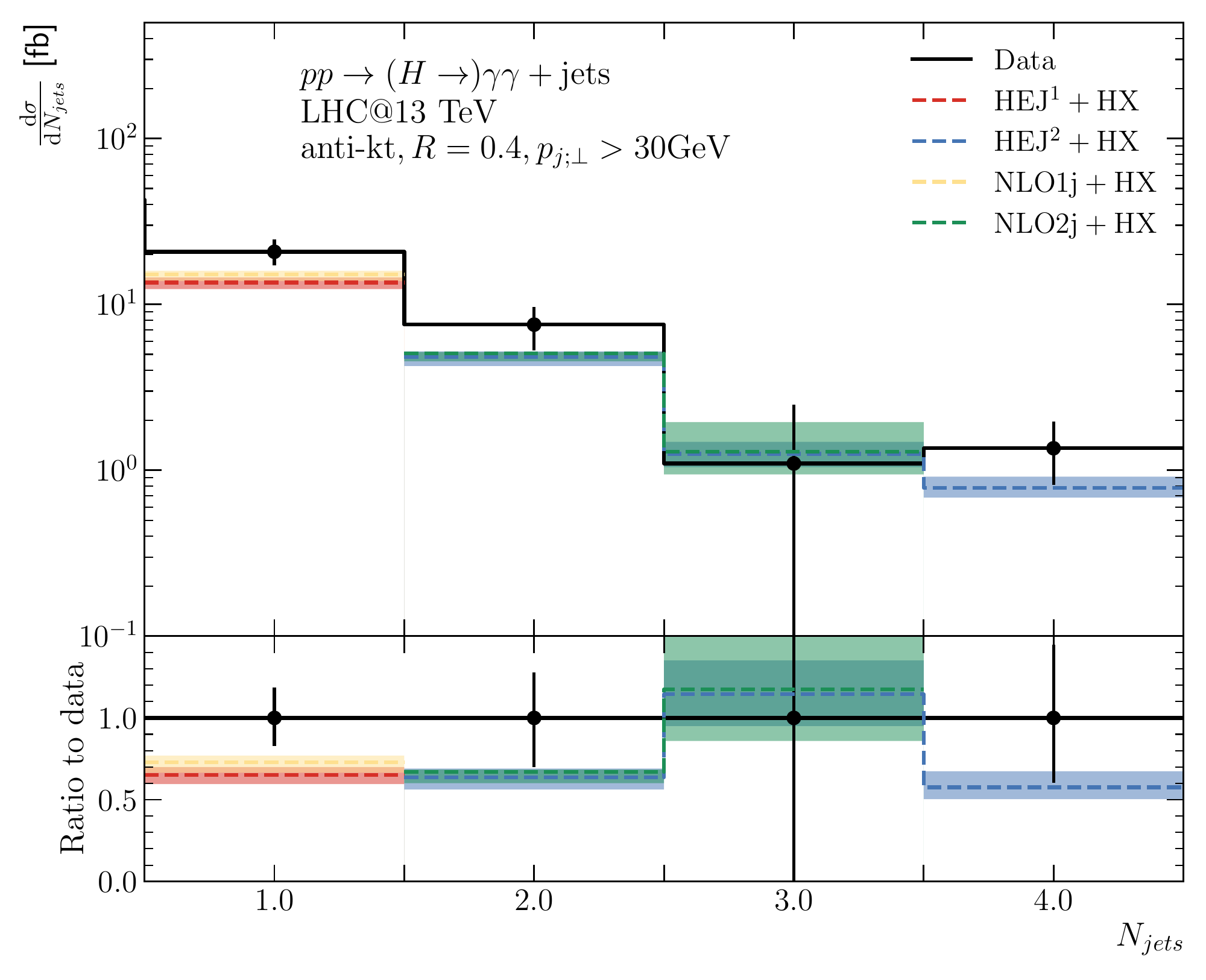}
            \caption[]%
            {}
            \label{fig:13TeV_CS_Njets}
        \end{subfigure}
        \hfill
        \begin{subfigure}[b]{0.475\textwidth}
            \centering
            \includegraphics[width=\textwidth]{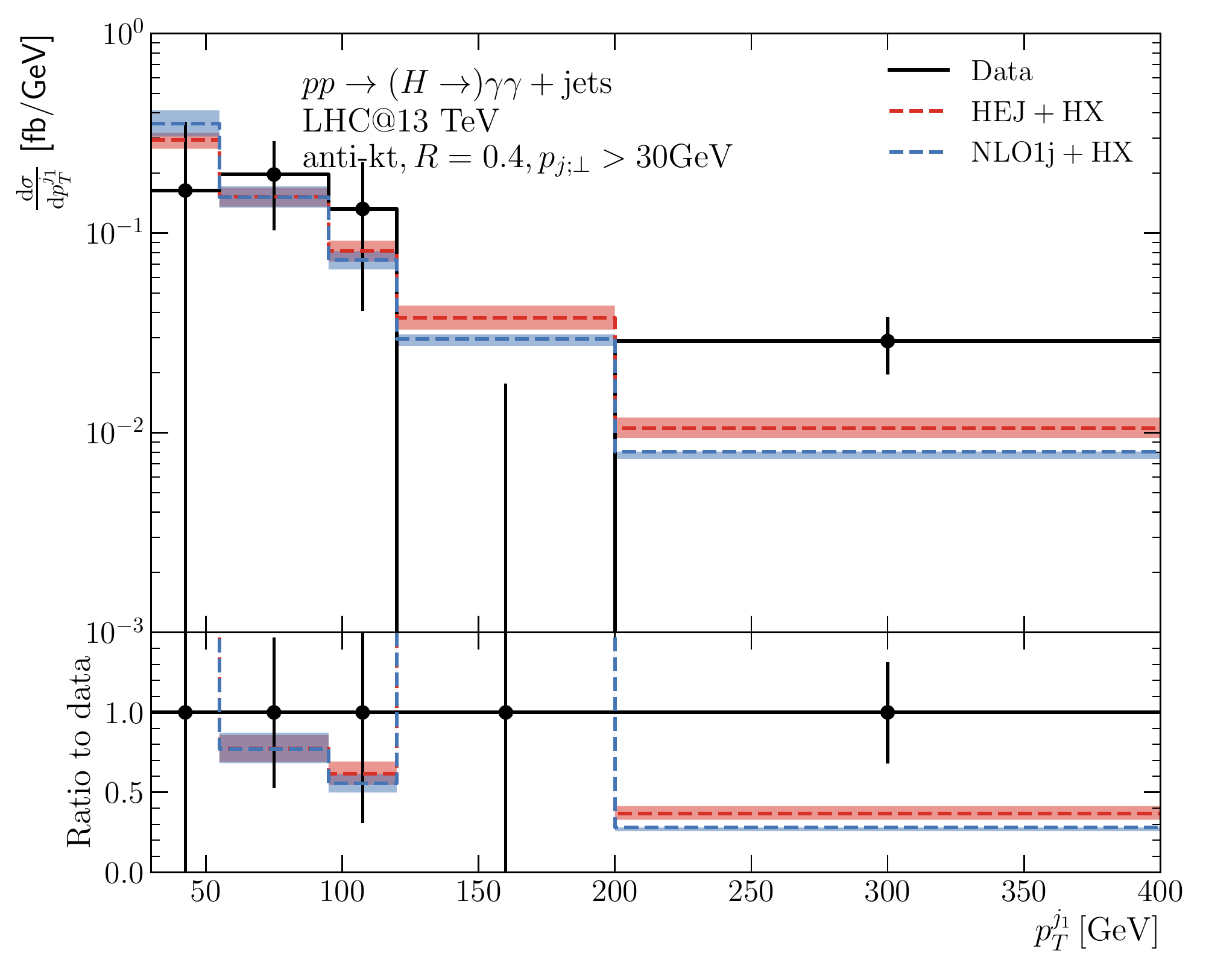}
           \caption[]%
           {}
            \label{fig:13TeV_CS_pTj1}
        \end{subfigure}
        \caption[ ]
        {\small (a) The exclusive number of jets compared to CMS data from~\cite{CMS:2022wpo}, and (b) the
          transverse momentum distribution of the leading jet compared to CMS
          data~\cite{CMS:2018ctp}. Both analyses employ the cuts
          described in table~\ref{table:13TeV-baseline}.  The ``HX'' component
          is extracted from those publications.}
        \label{fig:13TeV_Njets_pTj1}
\end{figure}

Figure~\ref{fig:13TeV_CS_Njets} shows the exclusive number of jets where the
1-jet and 2-jet \HEJ predictions are rescaled as described in
equation~\eqref{HEJNLONJ}. The fixed-order predictions are limited to 2 jets at
NLO and 3 jets at LO, whereas HEJ allows us to make predictions for the $\ge4$-jet
bin and reasonable agreement is achieved throughout.

In figure~\ref{fig:13TeV_CS_pTj1}, the transverse momentum of the first jet is
shown. We have compared to data from \cite{CMS:2018ctp} here rather than
\cite{CMS:2022wpo} as it covers a larger range.  The discrepancy between NLO
and \HEJ predictions as the transverse momentum increases is due to the
resummation procedure, and has also been observed in $W$+jets processes (see
ref.~\cite{Andersen:2020yax}). The effect would be even more significant for
greater values of $p_T$, however the collected data does not probe this region
of phase-space.  We have previously observed that a similar harder
$p_T$-spectrum seen in $H+\ge2j$ processes in \HEJ leads to a greater
sensitivity to the effects of using finite top and bottom quark
masses~\cite{Andersen:2018kjg}.

\begin{figure}[H]
        \begin{subfigure}[b]{0.475\textwidth}
            \centering
            \includegraphics[width=\textwidth]{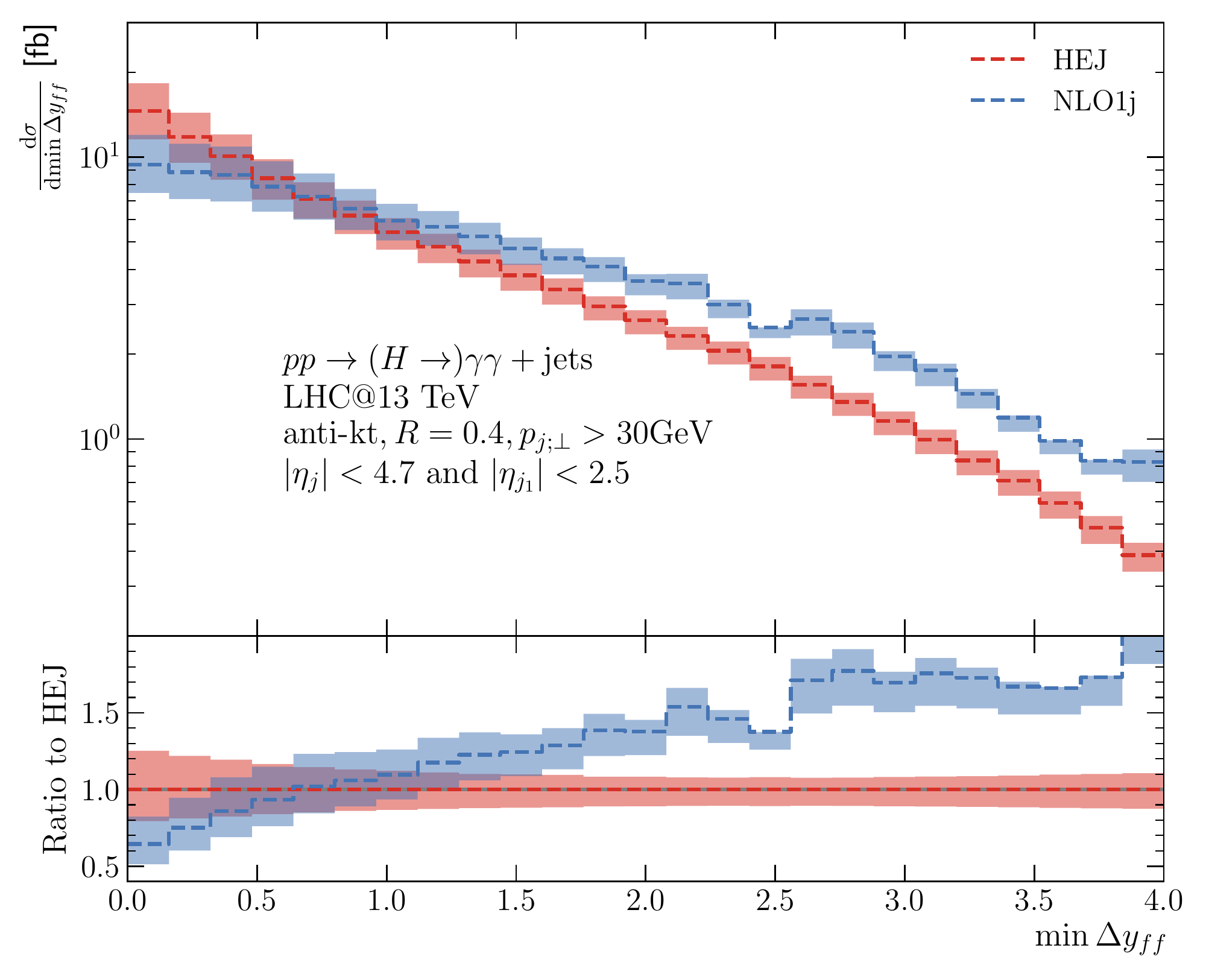}
            \caption[]%
            {}
            \label{fig:13TeV_CS_minDyany}
        \end{subfigure}
         \hfill
        \begin{subfigure}[b]{0.475\textwidth}
            \centering
            \includegraphics[width=\textwidth]{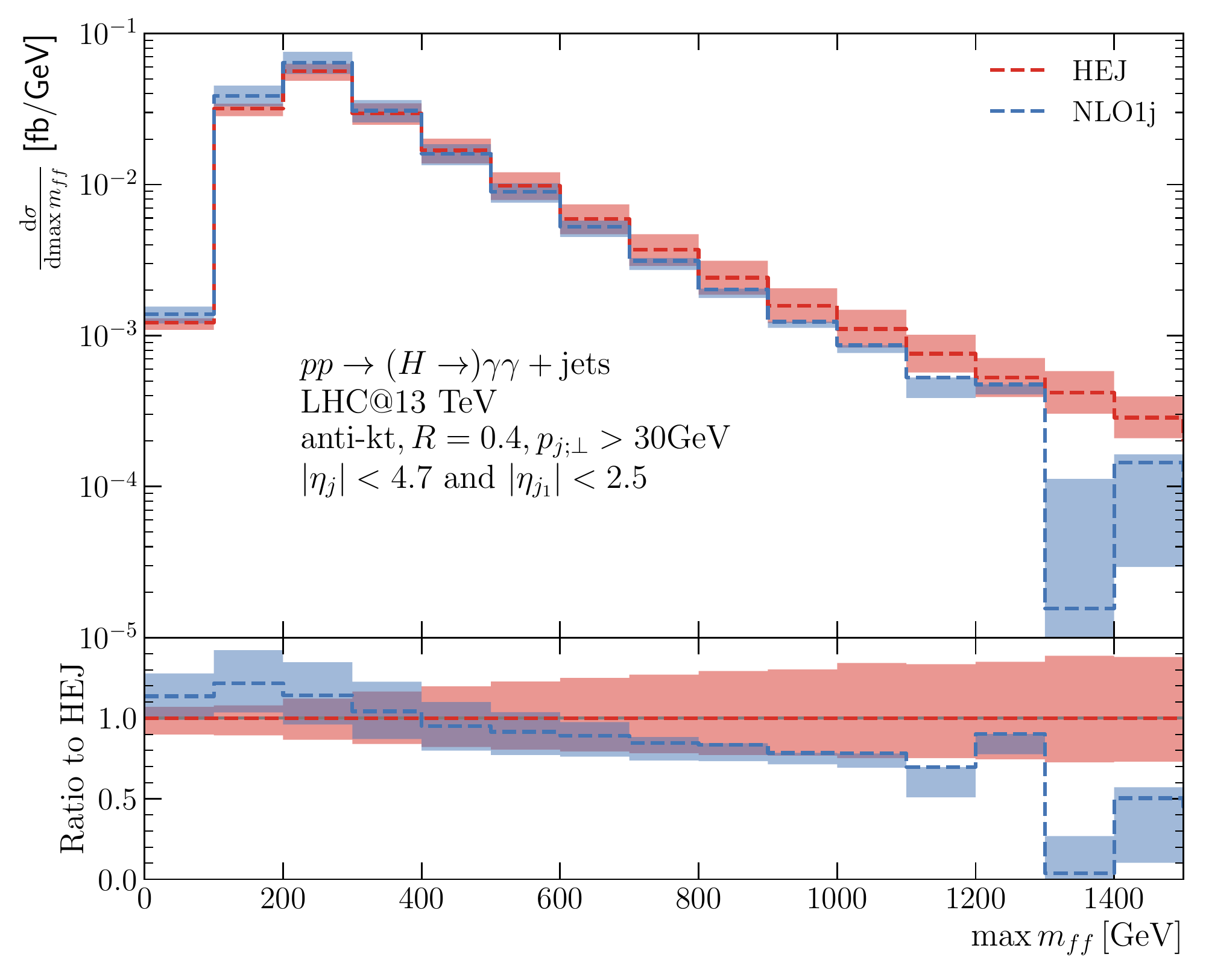}
            \caption[]%
            {}
            \label{fig:13TeV_CS_maxinvmassany}
        \end{subfigure}
        \caption[ ]
        {\small High-energy sensitive inclusive 1-jet
          distributions: (a) the minimum rapidity
          separation between any two outgoing particles (Higgs boson or
          jets) and (b) the maximum invariant mass
          between any two outgoing particles (Higgs or jets). \HEJ results are
          rescaled by the inclusive cross section ratio $\sigma_{\text{NLO1J}}/\sigma_{\text{\HEJ1J}}$.}
        \label{fig:13TeV_minDyany}
\end{figure}

The minimum rapidity separation between any two particles in the final state is
shown in figure~\ref{fig:13TeV_CS_minDyany}. As the Higgs boson is one of these
final states, this is a 1-jet observable, so the NLO 1-jet predictions are shown
for comparison and the \HEJ predictions are scaled by the ratio of the NLO to
\HEJ inclusive
1-jet rates. This observable is very sensitive to high-energy logarithmic corrections, and as was observed in
previous studies (see ref.~\cite{Andersen:2018kjg}), the effect of the
resummation results in a significant lowering of the \HEJ prediction compared to
fixed-order, by as much as 50\% at large values.
Figure~\ref{fig:13TeV_CS_maxinvmassany} shows the maximum invariant mass between
any two particles in the final state.  This is related to the high energy limit
where all pairwise invariant masses are taken to be large, but also includes
situations where two or more particles have a small invariant mass.  The impact
of the logarithmic corrections is not as strong here, and the fixed-order and
resummed predictions agree within uncertainties.

\subsection{Predictions for 8 TeV and Comparison to Data}
\label{sec:8TeV}

We now present predictions for an ATLAS analysis~\cite{ATLAS:2014yga} at a centre-of-mass energy of $\sqrt{\hat{s}} = 8$ TeV as implemented in Rivet\cite{Bierlich:2019rhm}. We list the relevant
experimental cuts used in this analysis in table~\ref{table:8TeV-baseline}, the complete list being available in
the experimental publication.  As in the experimental analysis, the jets are
reconstructed with the anti-$k_T$ algorithm with a radius parameter of $R=0.4$.
This study explored the inclusive and differential cross-sections for Higgs
boson production in the diphoton decay channel. For our purposes, we select the
observables which correspond to Higgs boson production plus at least one jet,
where our predictions are applicable.

\begin{table}[H]
\begin{center}
\begin{tabular}{|c|c|}
 \hline
Description & Baseline cuts  \\
 \hline
Photon transverse momentum & $p_T(\gamma) > 25 \text{ GeV}$  \\
Diphoton invariant mass & $105 \text{ GeV} < m_{\gamma\gamma} < 160 \text{ GeV} $ \\
Pseudo-rapidity of the photons & $|\eta_{\gamma} |<2.37$ excluding $1.37< |\eta_\gamma| < 1.56$\\
Ratio of harder photon $p_T$ to diphoton invariant mass & $p_T(\gamma_1)/m_{\gamma \gamma} > 0.35 $  \\
Ratio of softer photon $p_T$ to diphoton invariant mass & $p_T(\gamma_2)/m_{\gamma \gamma} > 0.25 $  \\
Photon isolation cut & $\text{Iso}^\gamma_\text{gen} < 14 \text{ GeV}$ \\
\hline
Jet transverse momentum & $p_T(j) > 30 \text{ GeV}$  \\
Jet rapidity & $|y_j| <4.4$  \\
\hline
\end{tabular}
\caption{Baseline cuts of the 8 TeV analysis, following the ATLAS analysis of~\cite{ATLAS:2014yga}. $\text{Iso}^\gamma_\text{gen}$ denotes the sum of transverse energies of stable particles in a cone of radius $\Delta R$ = 0.4 around each photon.}
\label{table:8TeV-baseline}
\end{center}
\end{table}

We divide our results into 1-jet observables, i.e.~containing at least one jet,
where the new components of \HEJ as detailed in section~\ref{sec:scaling_h1j}
can be tested, and 2-jet observables.  As in the previous
subsection, the experimental data points here include a non-GF contribution.  We
have extracted this ``HX'' component from~\cite{ATLAS:2014yga} where this was available.

\subsubsection{$H+\ge1j$}
\label{sec:h+ge1j}

\begin{figure}[H]
        \centering
        \begin{subfigure}[b]{0.475\textwidth}
            \centering
            \includegraphics[width=\textwidth]{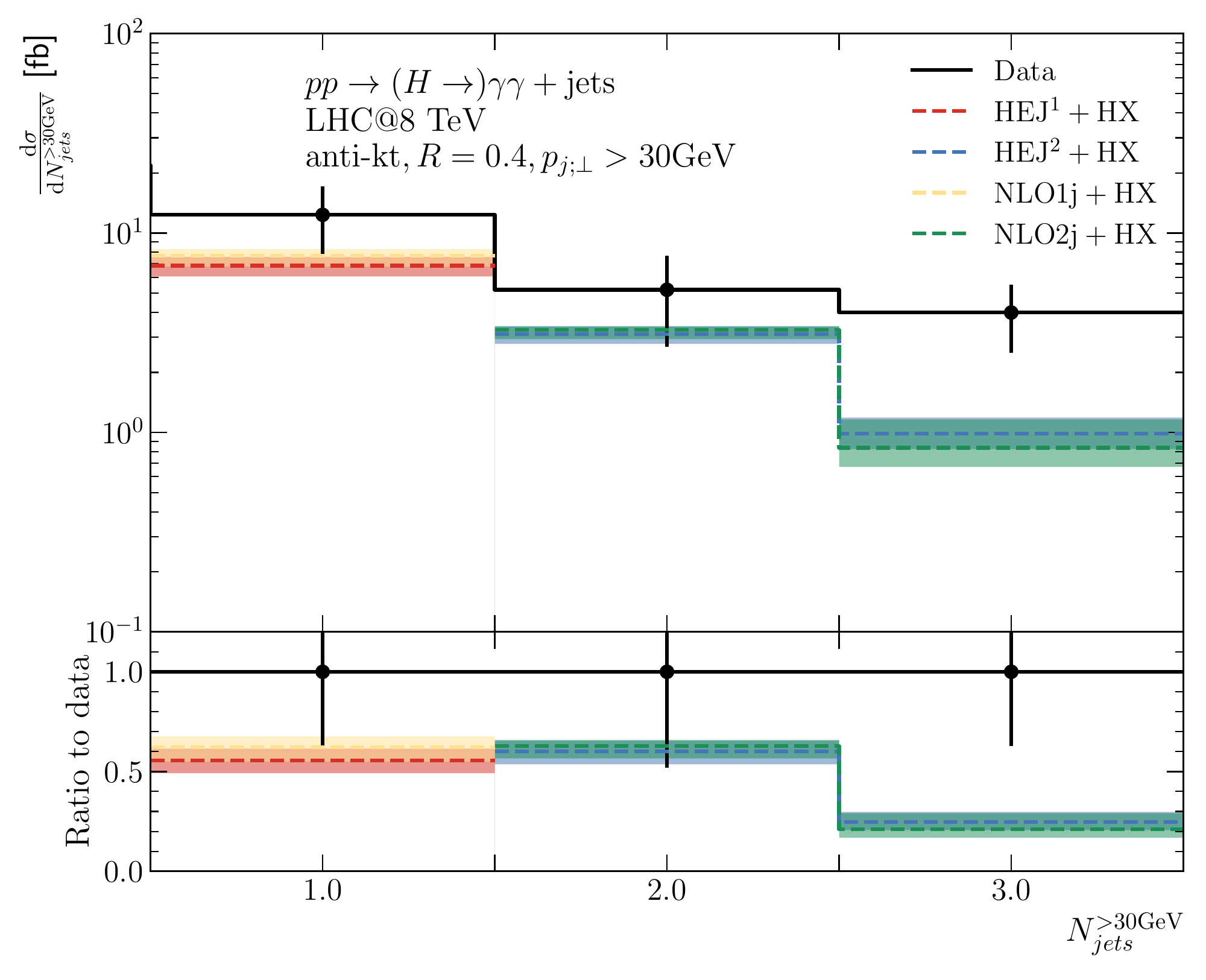}
            \caption[]%
            {{\small Number of jets}}
            \label{fig:8TeV_Njets30}
        \end{subfigure}
        \hfill
          \begin{subfigure}[b]{0.475\textwidth}
          \centering
          \includegraphics[width=\textwidth]{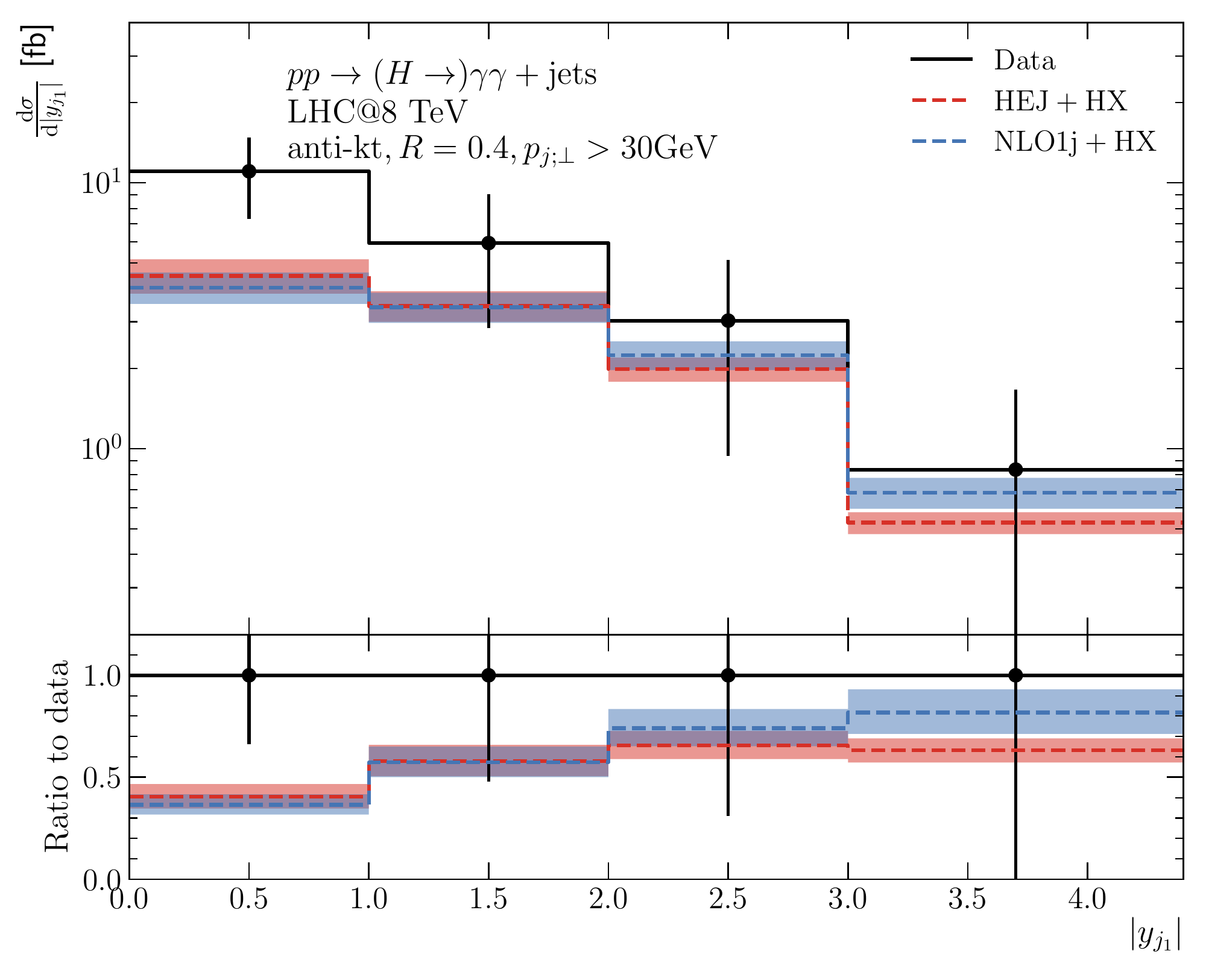}
          \caption[]%
          {{\small Leading jet rapidity}}
          \label{fig:8TeV_yj1}
        \end{subfigure}
        \vskip\baselineskip
        \begin{subfigure}[b]{0.475\textwidth}
            \centering
            \includegraphics[width=\textwidth]{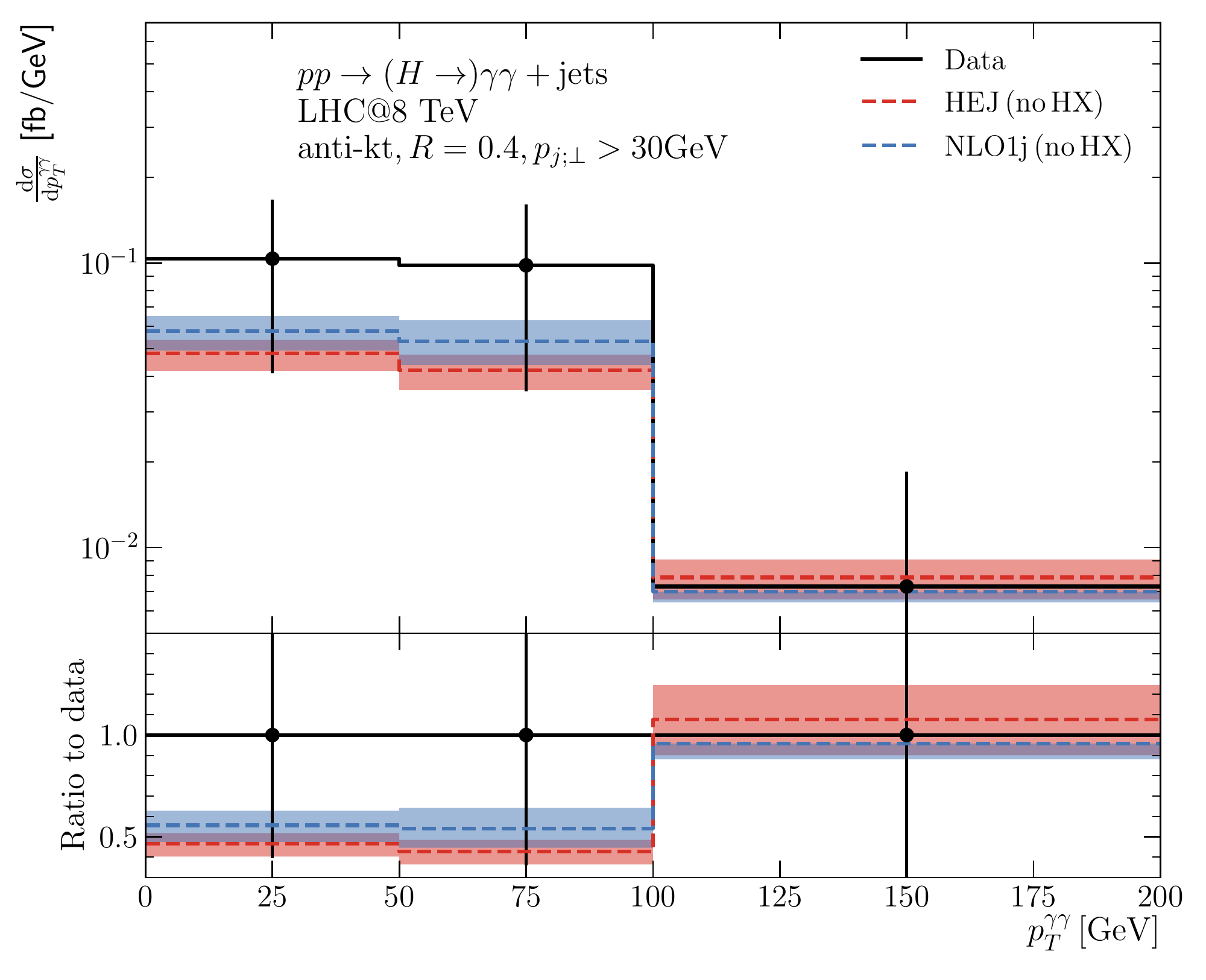}
            \caption[]%
            {{\small $p_T$ of the Higgs boson, $N_\text{jets} = 1$}}
            \label{fig:8TeV_pTgg1j}
        \end{subfigure}
        \hfill
        \begin{subfigure}[b]{0.475\textwidth}
            \centering
            \includegraphics[width=\textwidth]{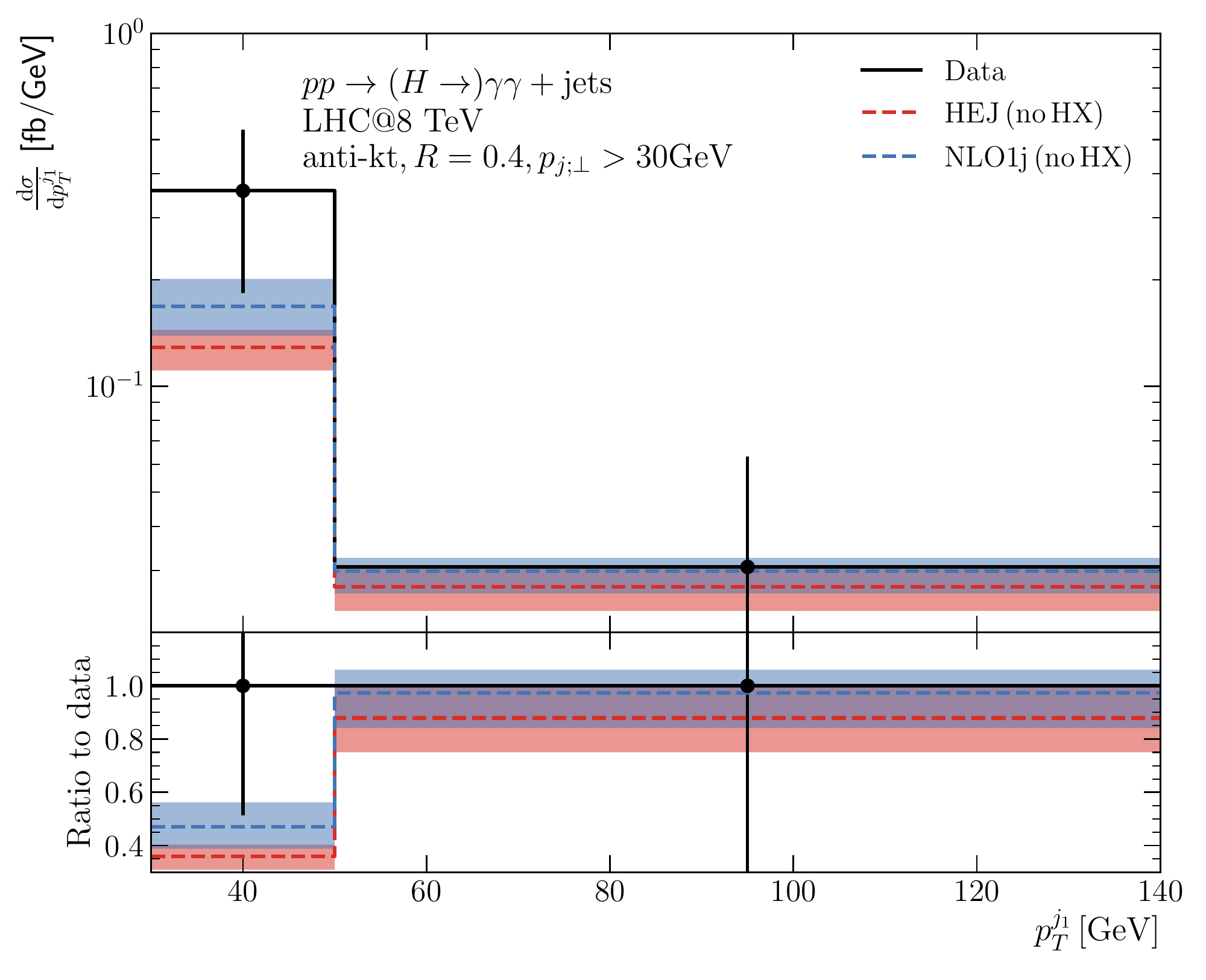}
            \caption[]%
            {{\small Leading jet $p_T$, $N_\text{jets} = 1$}}
            \label{fig:8TeV_pTj1}
        \end{subfigure}
        \caption[ ]
        {\small (\ref{fig:8TeV_Njets30}): Number of jets
          (exclusive). (\ref{fig:8TeV_yj1}): Leading jet
          rapidity. (\ref{fig:8TeV_pTgg1j}): Higgs boson transverse momentum in
          the 1-jet bin. (\ref{fig:8TeV_pTj1}): Leading jet transverse momentum
          in the 1-jet bin. The 1-jet \HEJ predictions are rescaled by the
          inclusive cross section ratio
          $\sigma_{\text{NLO1J}}/\sigma_{\text{\HEJ1J}}$ while the \HEJ
          predictions of the 2 and 3-jet bins of~(\ref{fig:8TeV_Njets30}) are
          rescaled by $\sigma_{\text{NLO2J}}/\sigma_{\text{\HEJ2J}}$. In
          (\ref{fig:8TeV_Njets30}) and (\ref{fig:8TeV_yj1}), the ``HX''
        component is extracted from~\cite{ATLAS:2014yga}; this was not available
      for (\ref{fig:8TeV_pTgg1j}) and (\ref{fig:8TeV_pTj1}).}
        \label{fig:8TeV_1jet}
    \end{figure}

In figure~\ref{fig:8TeV_Njets30}, we show the exclusive number of jets. As was
evidenced at 13 TeV, the differences between fixed-order and resummed predictions
are limited after the inclusive cross sections are rescaled.
The NLO and \HEJ predictions for
the 1- and 2-jet rates are
such that the bands for the theoretical scale variance and data uncertainty bands overlap.  The predictions in the $\ge3$-jet bin remain
slightly below data.

In figure~\ref{fig:8TeV_yj1}, the rapidity of the leading jet is displayed: the
discrepancy between the fixed-order and the resummed  predictions increases as
the rapidity of the jet attains large values. This is a High-Energy effect as opposed to a finite quark mass effect. Indeed, the corrections in $\hat{s}/t$ are particularly sizeable in this region of phase-space, and previous studies (see. ref~\cite{Andersen:2018kjg}) showed little dependence on the inclusion of the finite quark mass effects on this observable. However, this is not the case for the transverse momentum of the Higgs boson of figure~\ref{fig:8TeV_pTgg1j}. The finite quark mass effects and the resummation lead to a hardening of the high-$p_T$ tail of the Higgs boson, which would be even more dramatic had that region been probed.
Due to the probed phase-space region of $p_T  < 140 \, \text{GeV}$, \HEJ and fixed-order predictions for the hardest jet transverse momentum of figure~\ref{fig:8TeV_pTj1} remain close together and difficult to disentangle.

\begin{figure}[H]
                \begin{subfigure}[b]{0.475\textwidth}
                    \centering
                    \includegraphics[width=\textwidth]{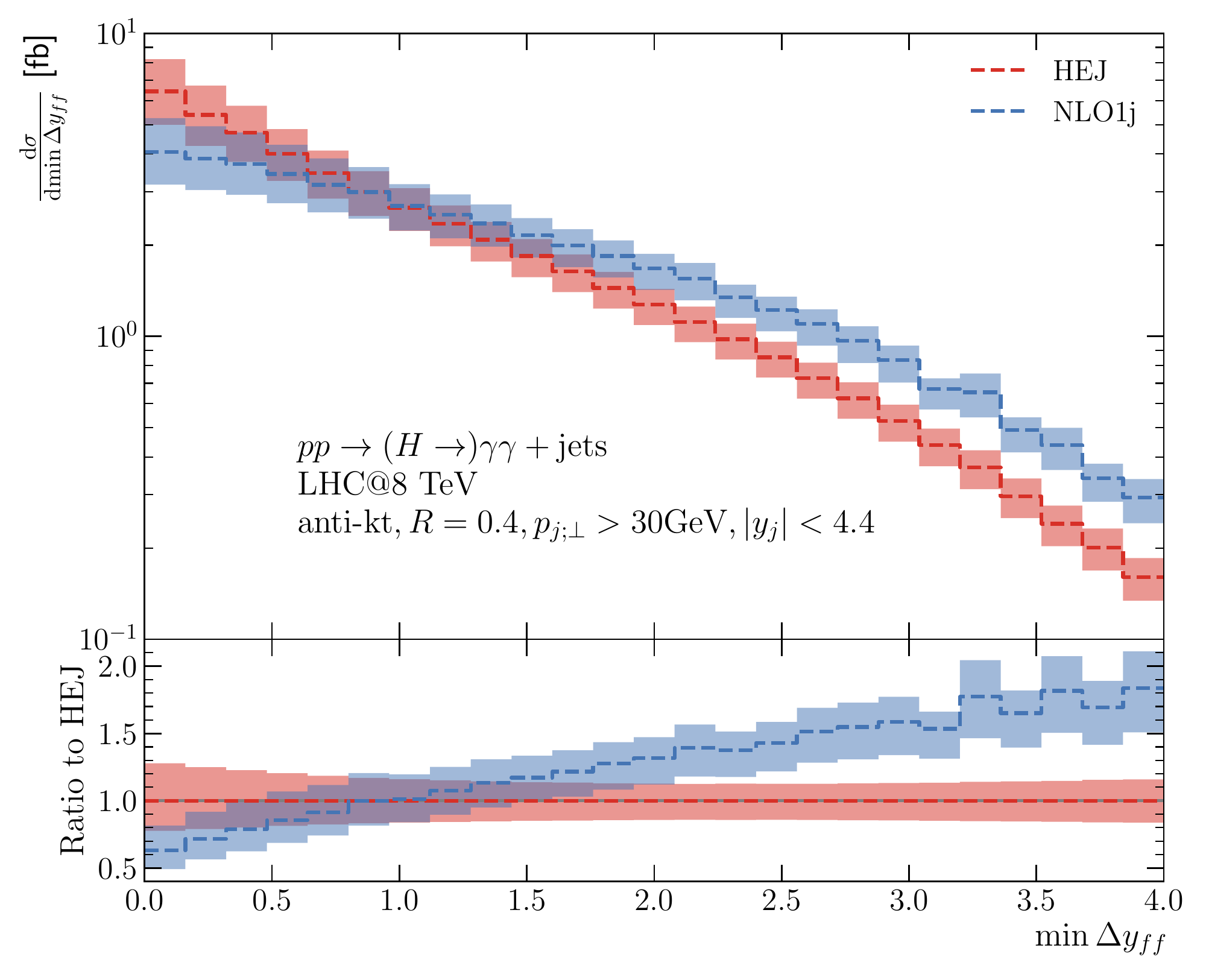}
                    \caption[]%
                    {{\small $\text{min }   \Delta y_{ff}$}}
                    \label{fig:8TeV_minDyany}
                \end{subfigure}
                \hfill
                \begin{subfigure}[b]{0.475\textwidth}
                    \centering
                    \includegraphics[width=\textwidth]{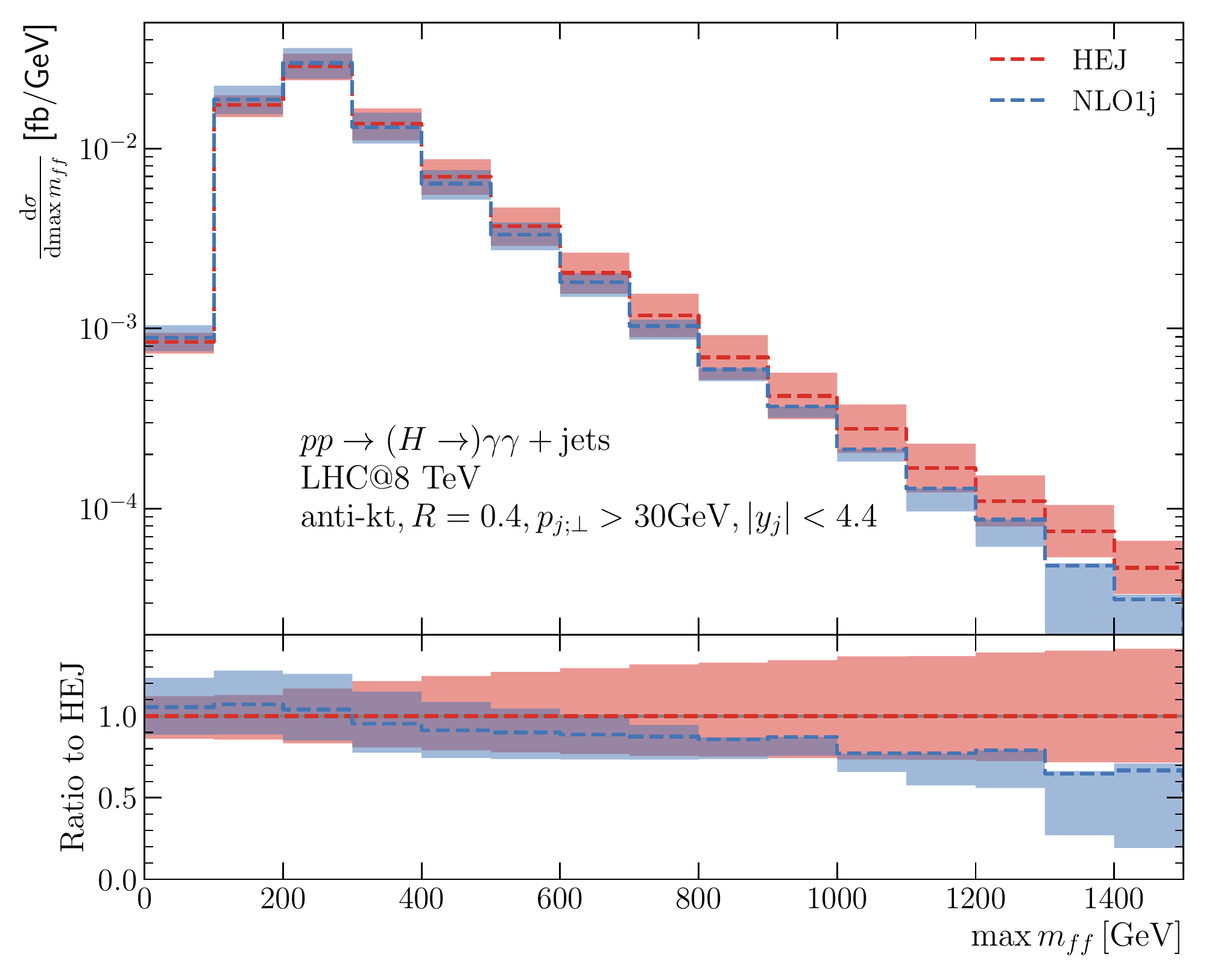}
                    \caption[]%
                    {{\small $\text{max }   m_{ff}$ }}
                    \label{fig:8TeV_maxinvmassany}
                \end{subfigure}
                \caption[ ]
                {\small High-energy sensitive 1-jet distributions. (\ref{fig:8TeV_minDyany}): minimum rapidity separation between any two outgoing particles (Higgs boson or jets). (\ref{fig:8TeV_maxinvmassany}): maximum invariant mass between any two outgoing particles (Higgs boson or jets). \HEJ results are rescaled by the inclusive cross section ratio $\sigma_{\text{NLO1J}}/\sigma_{\text{\HEJ1J}}$.}
                 \label{fig:8TeV_1j_more_no_data}
            \end{figure}

The High-Energy sensitive observables of figure~\ref{fig:8TeV_1j_more_no_data}
behave in a similar fashion to those at 13~TeV (figure~\ref{fig:13TeV_minDyany})
for the reasons explained in section~\ref{sec:13TeV}.  Here, there is nearly a
factor of two difference between NLO and \HEJ at large values of min $\Delta y_{ff}$.

\subsubsection{$H+\ge2j$}
\label{sec:h+ge2j}

\begin{figure}[H]
        \centering
        \begin{subfigure}[b]{0.475\textwidth}
            \centering
            \includegraphics[width=\textwidth]{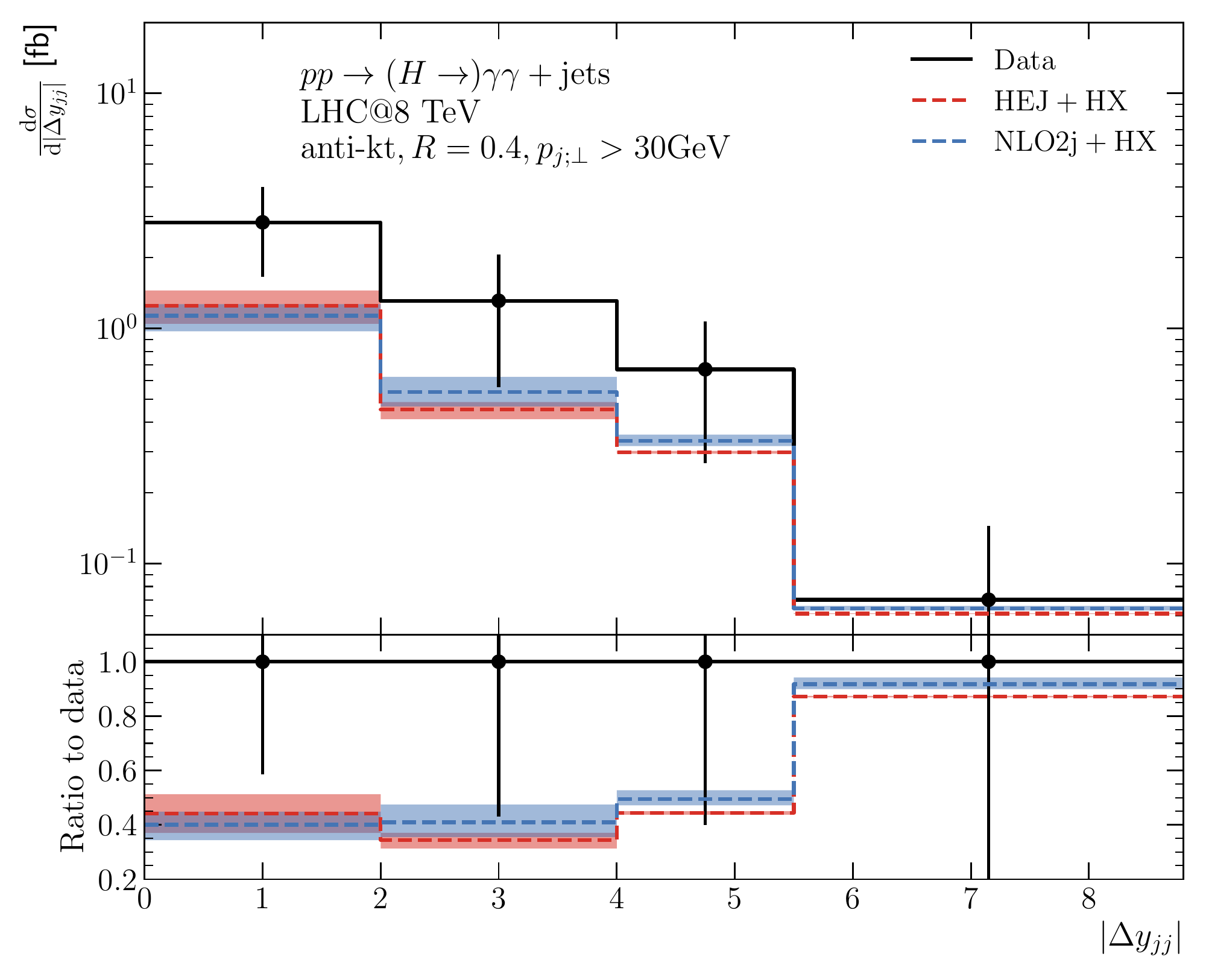}
            \caption[]%
            {{\small Dijet rapidity separation}}
            \label{fig:8TeV_dyjj}
        \end{subfigure}
        \hfill
        \begin{subfigure}[b]{0.475\textwidth}
            \centering
            \includegraphics[width=\textwidth]{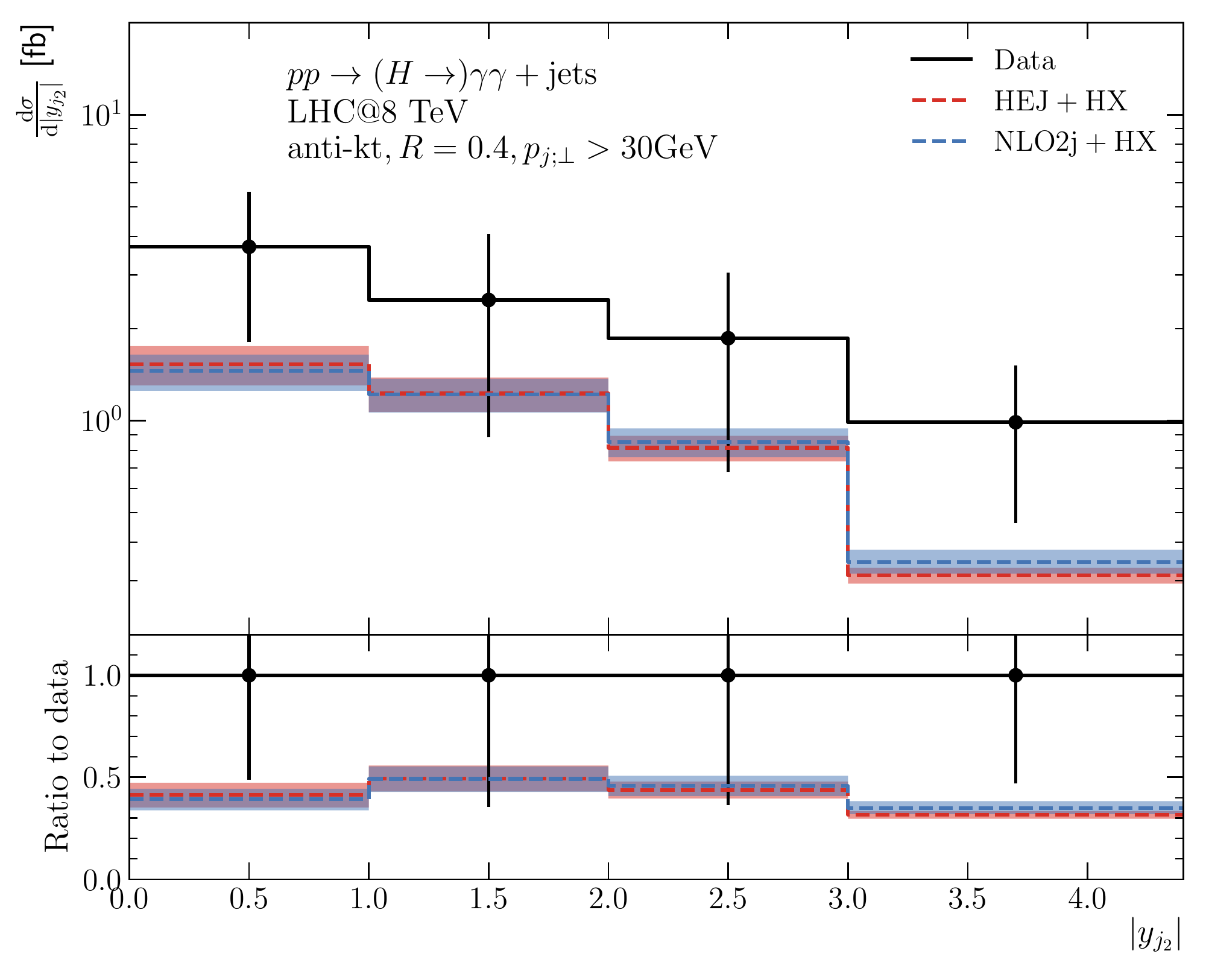}
            \caption[]%
            {{\small Subleading jet rapidity}}
            \label{fig:8TeV_yj2}
        \end{subfigure}
        \vskip\baselineskip
            \begin{subfigure}[b]{0.475\textwidth}
                \centering
                \includegraphics[width=\textwidth]{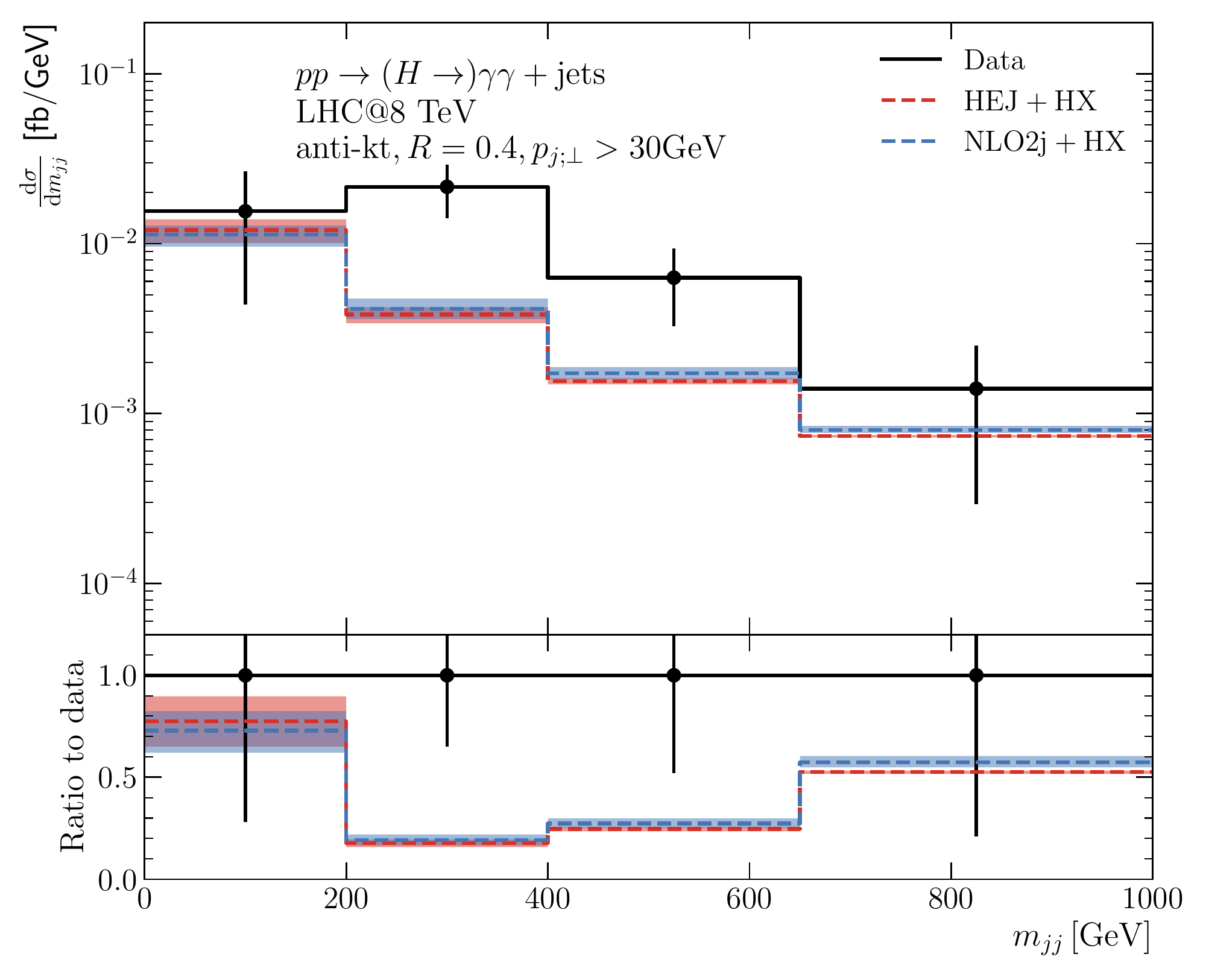}
                \caption[]%
                {{\small Invariant dijet mass}}
                \label{fig:8TeV_mjj}
            \end{subfigure}
        \hfill
        \begin{subfigure}[b]{0.475\textwidth}
            \centering
            \includegraphics[width=\textwidth]{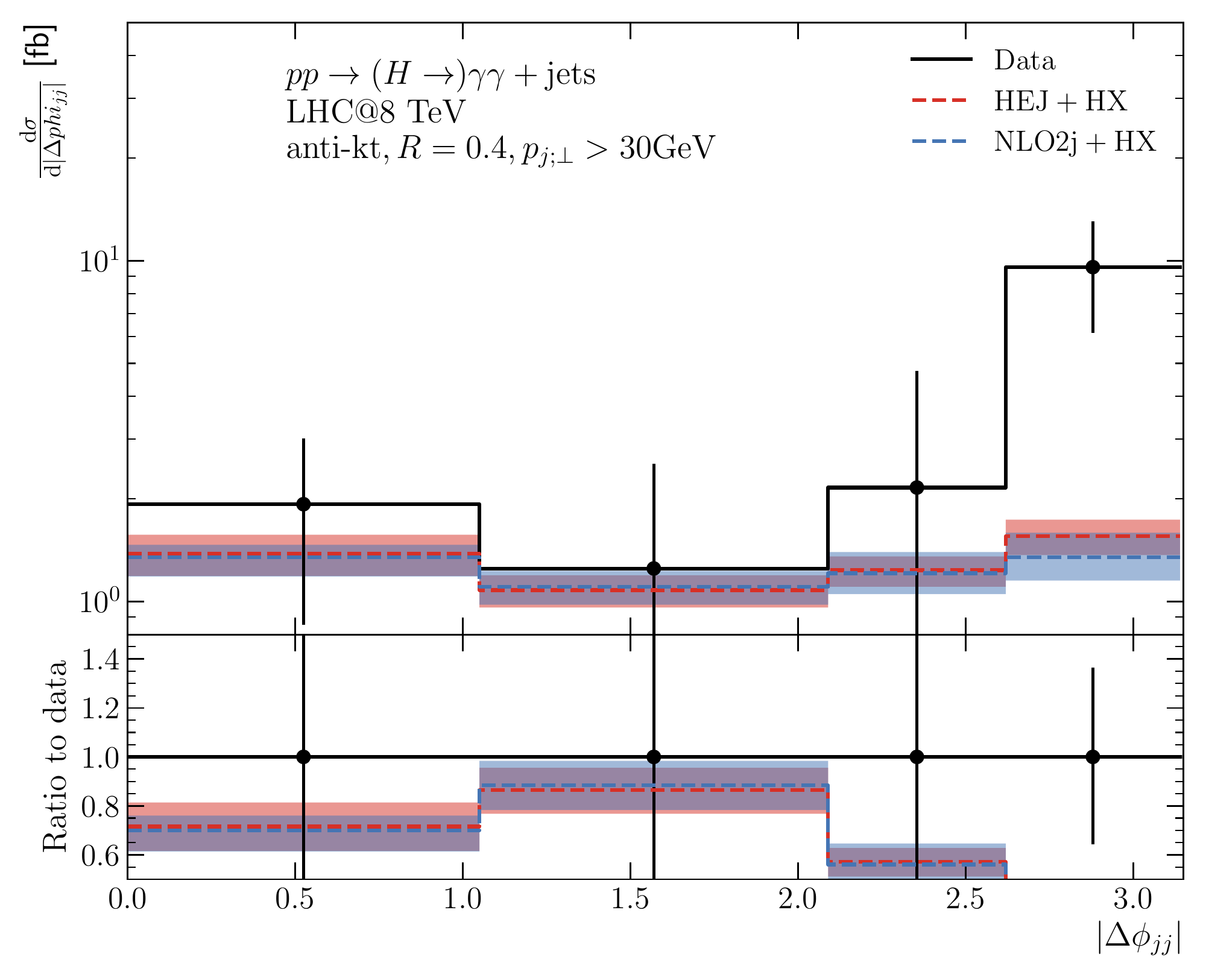}
            \caption[]%
            {{\small $\Delta \phi$ between the leading 2 jets}}
            \label{fig:8TeV_dphijj}
        \end{subfigure}
        \caption[ ]
        {\small
          (\ref{fig:8TeV_dyjj}): Dijet rapidity separation.
          (\ref{fig:8TeV_yj2}): Subleading jet rapidity.
          (\ref{fig:8TeV_mjj}): dijet invariant mass.
          (\ref{fig:8TeV_dphijj}): Azimuthal angle difference between the leading 2 jets.
          All 2-jet \HEJ predictions are rescaled by the inclusive cross section
          ratio $\sigma_{\text{NLO2J}}/\sigma_{\text{\HEJ2J}}$.  The ``HX''
          component is extracted from~\cite{ATLAS:2014yga}.
        }
        \label{fig:8TeV_2j}
    \end{figure}

\begin{figure}[H]
            \centering
            \begin{subfigure}[b]{0.475\textwidth}
              \centering
              \includegraphics[width=\textwidth]{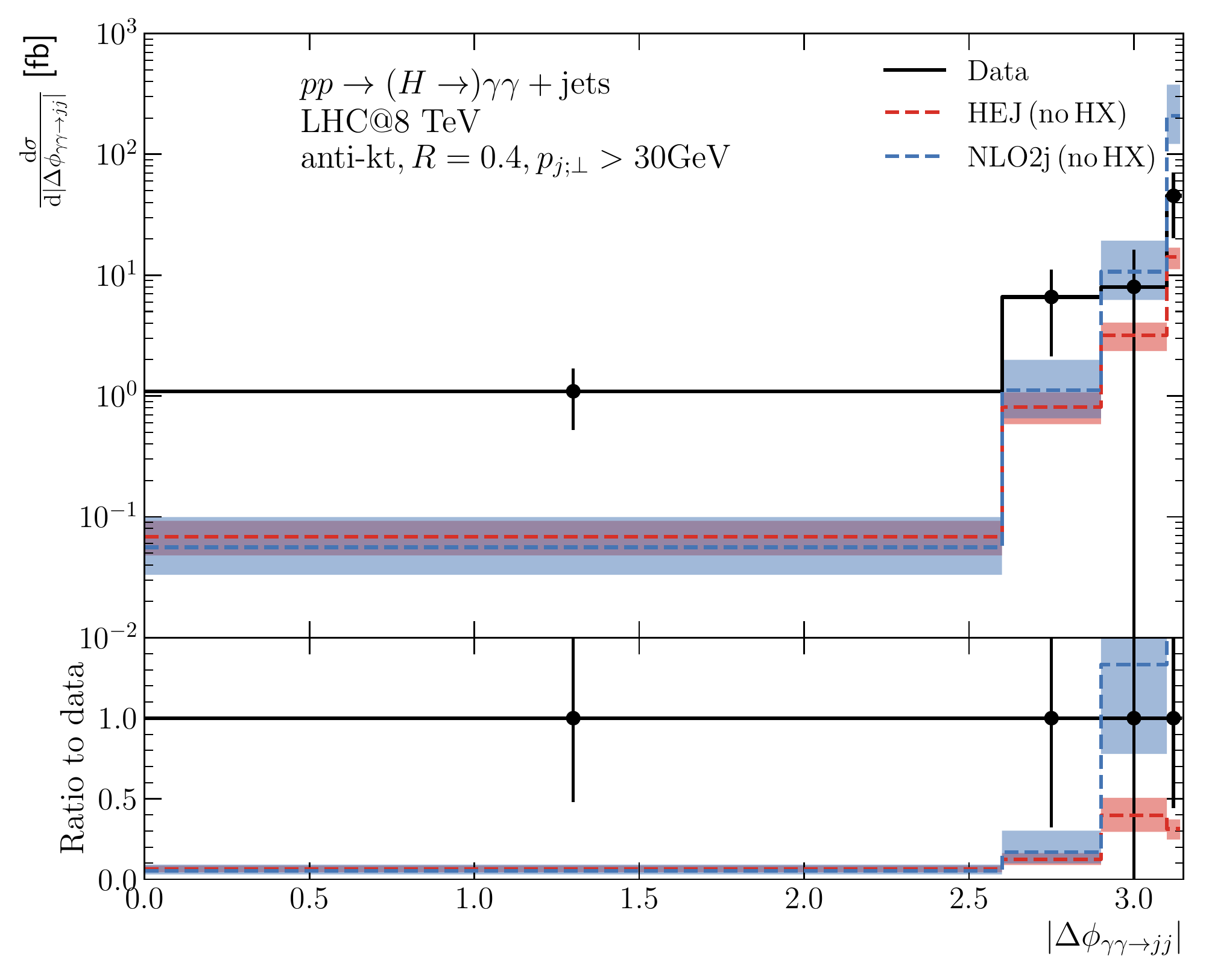}
              \caption[]%
              {{\small $\Delta \phi$ between dijet and diphoton systems}}
              \label{fig:8TeV_dphiggjj}
            \end{subfigure}
            \hfill
            \begin{subfigure}[b]{0.475\textwidth}
                \centering
                \includegraphics[width=\textwidth]{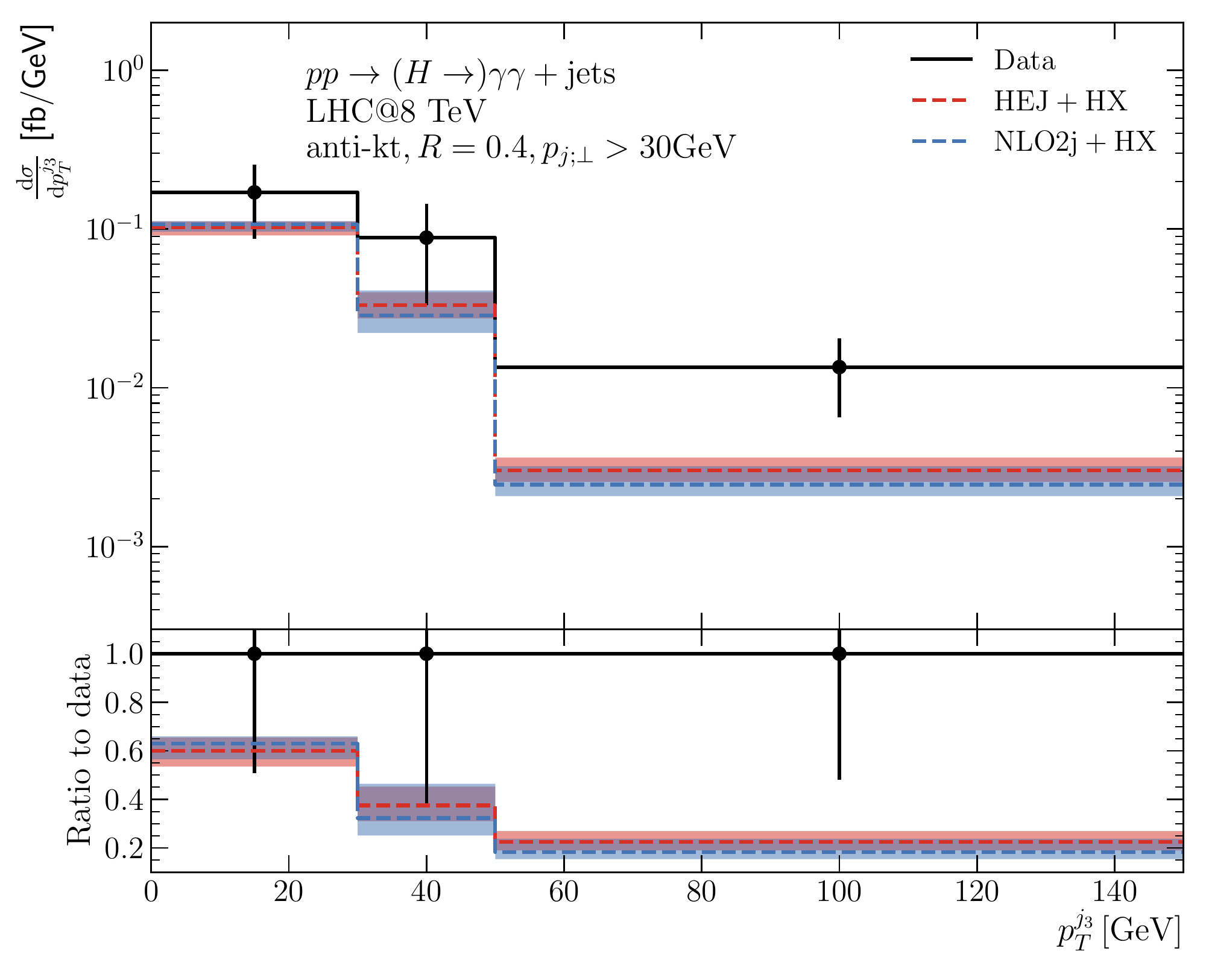}
                \caption[]%
                {{\small Third-leading jet $p_T$}}
                \label{fig:8TeV_pTj3}
            \end{subfigure}
            \vskip\baselineskip
            \begin{subfigure}[b]{0.475\textwidth}
                \centering
                \includegraphics[width=\textwidth]{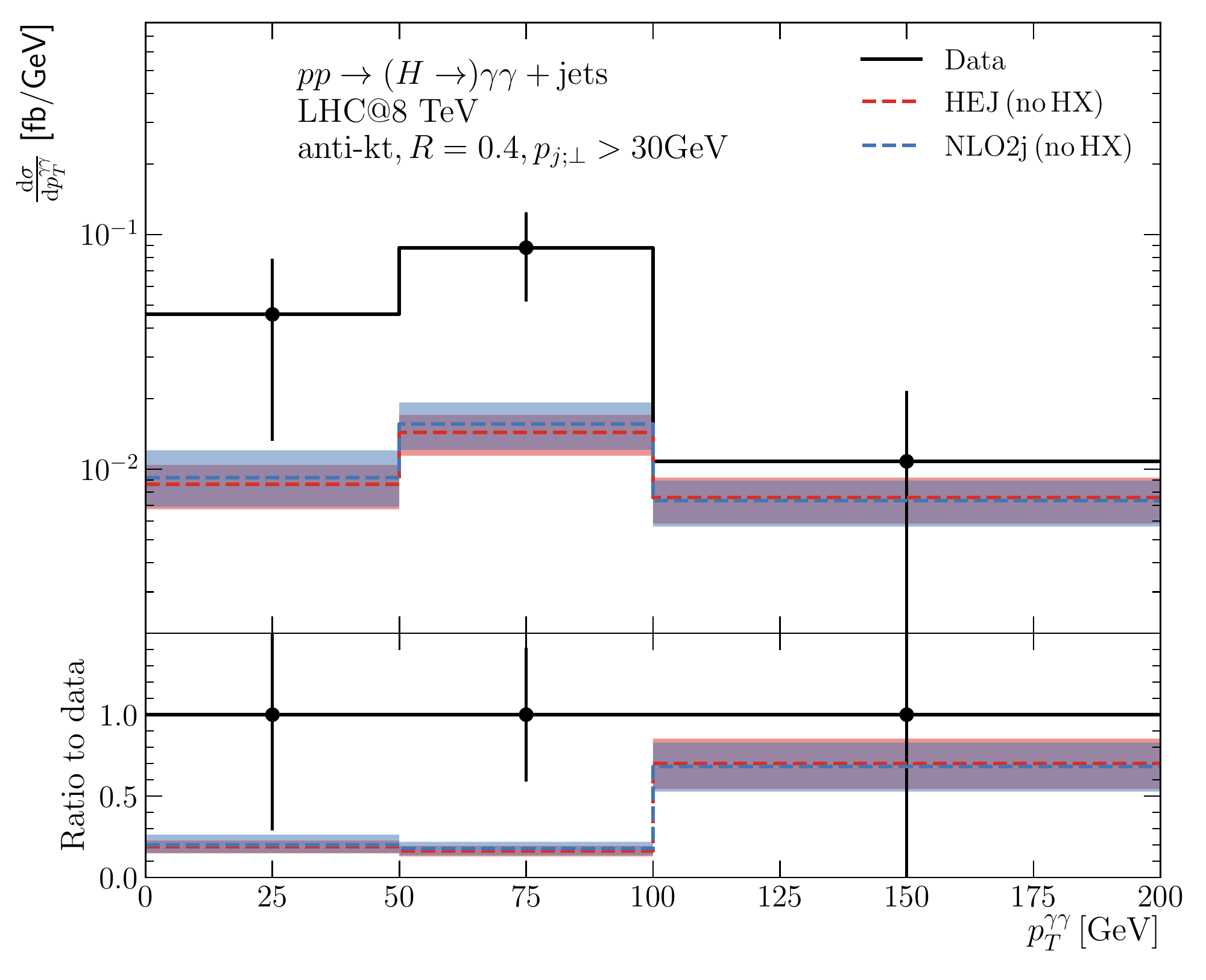}
                \caption[]%
                {{\small Higgs $p_T$ with $N_\text{jets}=2$}}
                \label{fig:8TeV_pTgg2j}
            \end{subfigure}
            \hfill
            \begin{subfigure}[b]{0.475\textwidth}
                \centering
                \includegraphics[width=\textwidth]{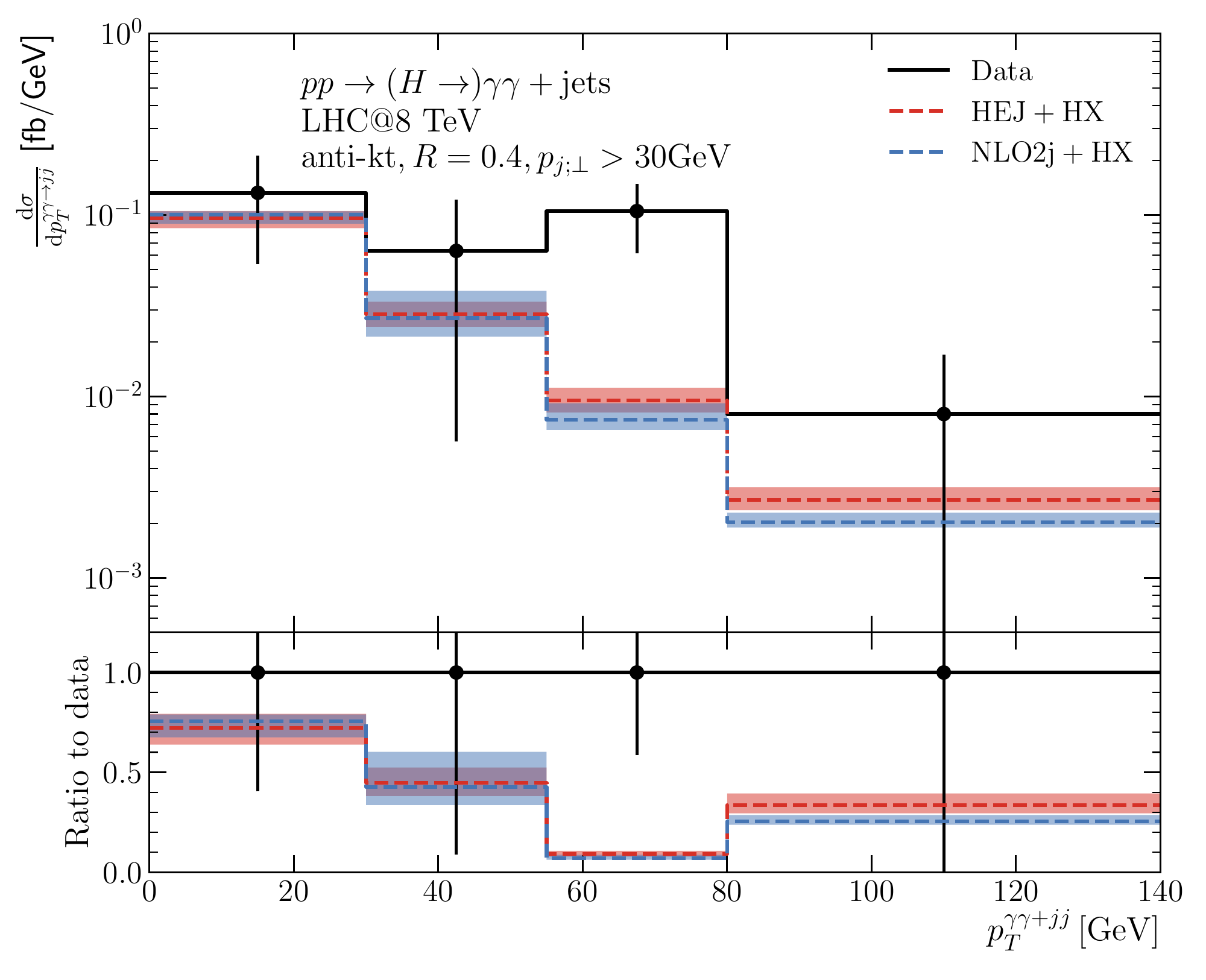}
                \caption[]%
                {{\small $p_T$ of diphoton-dijet system}}
                \label{fig:8TeV_pTggjj}
            \end{subfigure}
            \caption[ ]
            {\small
              (\ref{fig:8TeV_dphiggjj}): Azimuthal angle difference between dijet and diphoton objects.
              (\ref{fig:8TeV_pTj3}): Transverse momentum of the third-leading jet.
              (\ref{fig:8TeV_pTgg2j}): Transverse momentum of the Higgs boson in the $\geq$ 2-jet bin.
              (\ref{fig:8TeV_pTggjj}): Transverse momentum of the Higgs plus dijet object : $(p_H+p_{j_1} + p_{j_2})_\perp$.
              All 2 and 3-jet \HEJ predictions are rescaled by the inclusive
              cross section ratio
              $\sigma_{\text{NLO2J}}/\sigma_{\text{\HEJ2J}}$.  In
              (\ref{fig:8TeV_pTj3}) and (\ref{fig:8TeV_pTggjj}), the ``HX''
          component is extracted from~\cite{ATLAS:2014yga}; this was not
          available for (\ref{fig:8TeV_dphiggjj}) and (\ref{fig:8TeV_pTgg2j}).
            }
            \label{fig:8TeV_2j_more}
        \end{figure}

We now turn to a range of 2-jet observables, displayed in figures~\ref{fig:8TeV_2j}
and \ref{fig:8TeV_2j_more}.
Globally, the impact of the resummation on High-Energy sensitive observables is
to lower the predictions from fixed-order approaches, as can be seen in large
dijet rapidity separation in figure~\ref{fig:8TeV_dyjj}, large rapidity values
of the second hardest jet in figure~\ref{fig:8TeV_yj2} and at large dijet
invariant mass in figure~\ref{fig:8TeV_mjj}.  This can be seen more clearly
before the addition of the ``HX'' component, see figure~\ref{fig:8TeV_dyjj_nodata}
in appendix~\ref{sec:additional-plots-qcd}.  As expected, the resummation
procedure has little impact on the observables dependent on the azimuthal
degrees of freedom: the azimuthal angle difference between the leading two jets
of figure~\ref{fig:8TeV_dphijj} and the azimuthal angle difference between the
diphoton and the leading dijets systems depicted in
figure~\ref{fig:8TeV_dphiggjj}, expect perhaps at values close to $\pi$ (that is
when the systems are back-to-back).

As previously observed, the combination of the inclusion of corrections in
$\hat{s}/t$ and the finite quark mass effects tend to harden the tail of the
transverse momenta distributions compared to fixed order predictions. This is
apparent in the description of the third-leading jet transverse momentum of
figure~\ref{fig:8TeV_pTj3} (see figure~\ref{fig:8TeV_pTj3_nodata} for the shapes
of the pure QCD predictions), but also in the transverse momentum of the
diphoton-dijet system of figure~\ref{fig:8TeV_pTgg2j}. Although the Higgs
transverse momentum seems to be independent of the effect of the resummation, it
is conjectured that values of $p^{\gamma\gamma}_\perp$ above 200~GeV would lead to a
disparity between the two approaches.

%% file: Conclusions.tex
\section{Conclusions}
\label{sec:conclusions}

In this paper we have presented an alternative description of $pp\to H+\ge 1j$,
which is accurate to leading logarithms in $\hat s/p_T^2$ (LL).  We have
outlined the structure of a LL-accurate amplitude in the \HEJ formalism, and
described the calculation of the necessary new components in
section~\ref{sec:Hjets_he}.  One big advantage of the approach is that it
maintains full dependence on the finite top and bottom quark masses in the
couplings of the Higgs boson to gluons for \emph{any} number of jets, which
quickly exceeds the multiplicities currently calculated at even leading order.
The new pieces allow LL resummation in $\hat s/p_T^2$ to an inclusive 1-jet
process for the first time in the \HEJ framework.

We have then compared the resummed predictions to fixed-order predictions and
to LHC data in section~\ref{sec:comparisons-data}, and discussed the impact of
the logarithmic corrections.  We find the impact of the resummation is seen at
large jet transverse momenta.  The resummed results give a harder
$p_T$-spectrum compared to NLO, which in turn leads to a greater dependence on
finite quark masses in the coupling.  We also observe a large suppression
compared to NLO at large values of rapidity separation between all pairs of
final state particles (i.e.~between any two of the Higgs boson and jets).  This
can be as much of a factor of two and lies significantly outwith the uncertainty
bands on the two predictions.  Other
observables, e.g.~azimuthal angles, are less sensitive to these logarithmic corrections.

Looking forward to analyses of LHC Run 3 data, our results suggest that the
inclusion of finite quark
masses for higher jet multiplicities \emph{and} of logarithmic corrections in
$\hat s/p_T^2$ will be important in the comparison to data.

\section*{Acknowledgements}
We are grateful to the other members of the \HEJ collaboration for useful and
helpful discussions throughout this work.  We are pleased to acknowledge funding
from the UK Science and Technology Facilities Council (under grant number ST/T506047/1
for HH), the Royal Society and the ERC Starting Grant 715049 ``QCDforfuture''.  The
predictions presented in section~\ref{sec:comparisons-data} were produced using
resources from PhenoGrid which is part of the GridPP
Collaboration~\cite{gridpp2006,gridpp2009}.  AP acknowledges support by the
National Science Foundation under Grant No. PHY 2210161. For the purpose of open access,
the authors have applied a Creative Commons Attribution (CC BY) licence to any
Author Accepted Manuscript version arising from this submission.


%% file: Appendix.tex
\section{NLO reweighting factors}
\label{sec:appendix}

In table~\ref{table:reweighting-factors}, we give the value of the NLO
reweighting factors as described in equation~\eqref{HEJNLONJ} for both the 8~TeV~\cite{ATLAS:2014yga} and 13~TeV~\cite{CMS:2018ctp,CMS:2022wpo} analyses.

\begin{table}[h!]
\begin{center}
\begin{tabular}{lllllll}
\hline
Analysis  & \multicolumn{3}{l}{8 TeV}                                                & \multicolumn{3}{l}{13 TeV}                                               \\
\hline
Scale     & $\mu_F,\mu_R$               & $(\mu_F,\mu_R)/2$                &  $2(\mu_F,\mu_R)$                & $\mu_F,\mu_R$               & $(\mu_F,\mu_R)/2$                &  $2(\mu_F,\mu_R)$                 \\
\hline
1J factor & 1.87 & 1.54 & 2.15 & 1.59 & 1.30 & 1.84 \\
2J factor & 1.98 & 1.48 & 2.40 & 1.62 & 1.19 & 2.00 \\
\hline
\end{tabular}
\caption{NLO Reweighting factors with $\mu_F = \mu_R = \max(m_{12},m_H)$.}
\label{table:reweighting-factors}
\end{center}
\end{table}

In the 8 TeV analysis, the inclusive 1-jet and  2-jet cross-sections are calculated from the rapidity of the hardest and second hardest jet histograms respectively, figures~\ref{fig:8TeV_yj1} and~\ref{fig:8TeV_yj2}.

In the 13 TeV analysis, the inclusive 1-jet and 2-jet cross-sections are obtained from the appropriate $N_{\text{jets}}$ bins, figure~\ref{fig:13TeV_CS_Njets}. Note that for at least 2 jets, this plot requires central jets only, but it is valid to use it as it is the only plot we present for the 2-jet observables. If more inclusive cross-sections are considered, say from the rapidity of the second hardest jet histogram over all the experimental range, then the 2-jet reweighting factor would be further away from the 1-jet value (1.79 for the central scale instead of 1.62).


%% file: QCDPlots.tex
\section{Additional Plots of QCD Component}
\label{sec:additional-plots-qcd}

\begin{figure}[H]
        \centering
        \begin{subfigure}[b]{0.475\textwidth}
            \centering
            \includegraphics[width=\textwidth]{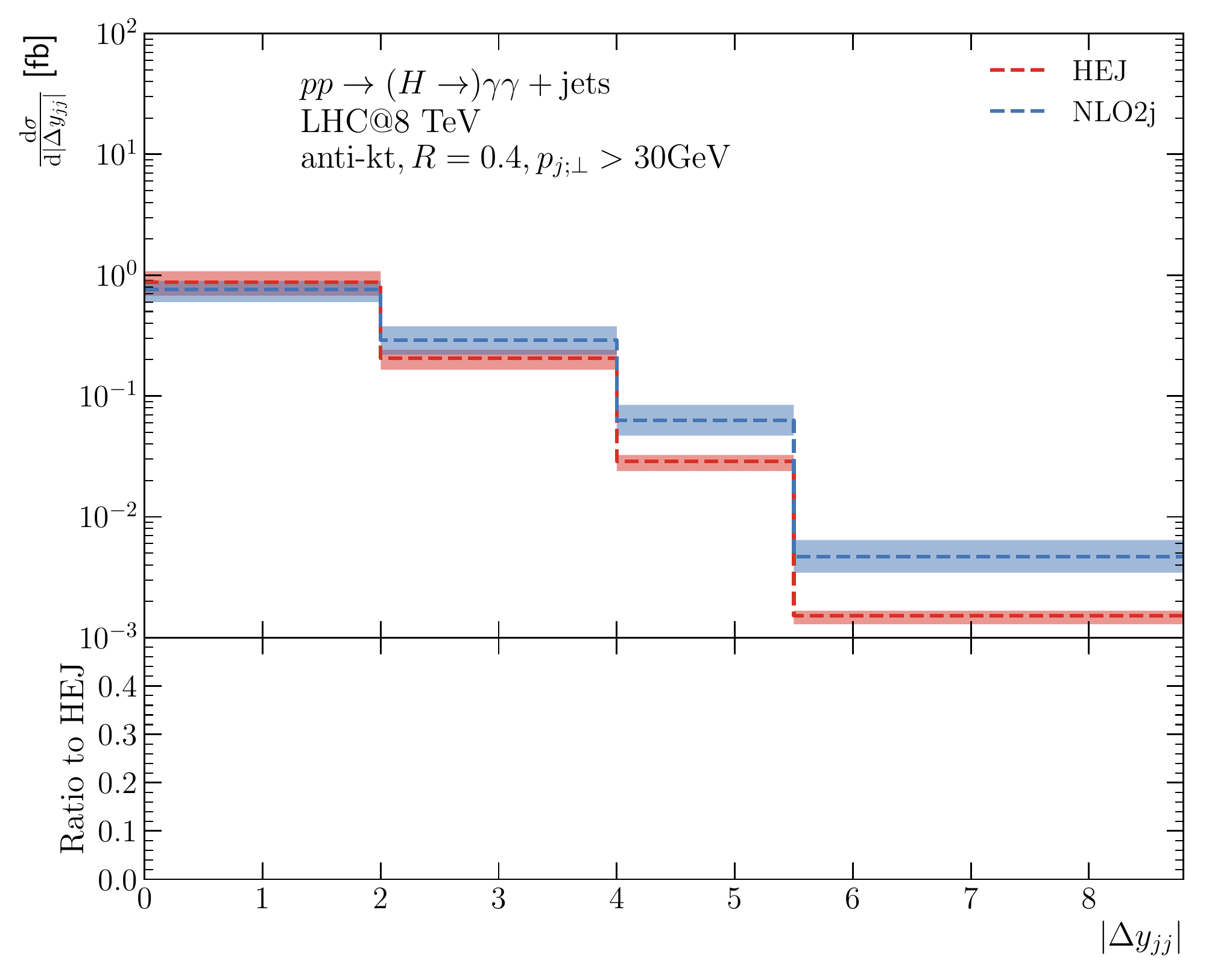}
            \caption[]%
            {{\small Dijet rapidity separation}}
            \label{fig:8TeV_dyjj_nodata}
        \end{subfigure}
        \hfill
        \begin{subfigure}[b]{0.475\textwidth}
            \centering
            \includegraphics[width=\textwidth]{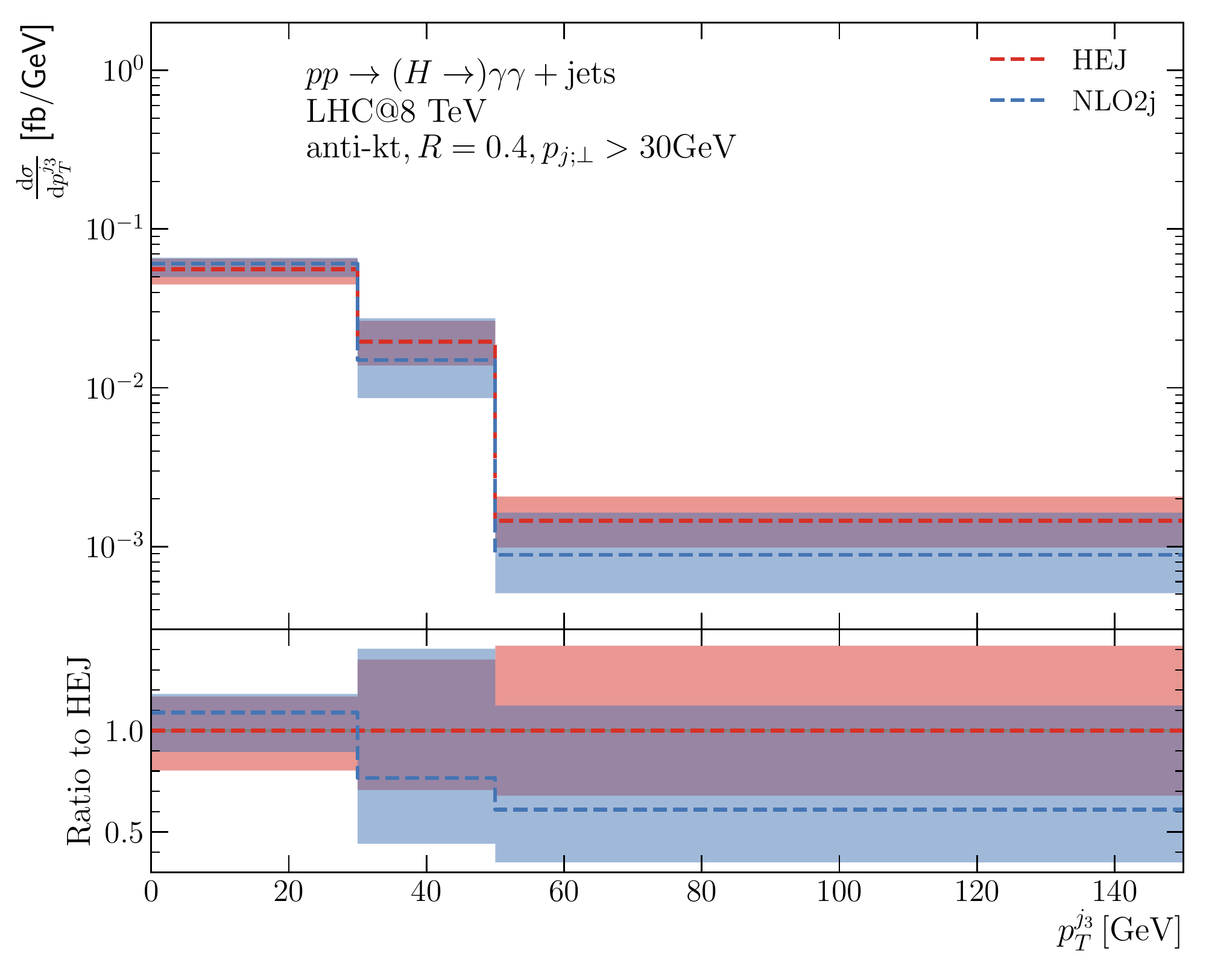}
            \caption[]%
            {{\small Subleading jet rapidity}}
            \label{fig:8TeV_pTj3_nodata}
        \end{subfigure}
        \caption[ ]
        {\small (\ref{fig:8TeV_dyjj_nodata}): Dijet rapidity separation.
          (\ref{fig:8TeV_dphijj}): Azimuthal angle difference between the
          leading 2 jets.  This shows the distributions from
          Figs.~\ref{fig:8TeV_dyjj} and \ref{fig:8TeV_pTj3}, where now we only show the QCD
          contribution.  As before, the 2-jet \HEJ predictions are rescaled by the
          inclusive cross section ratio
          $\sigma_{\text{NLO2J}}/\sigma_{\text{\HEJ2J}}$.  }
        \label{fig:8TeV_2j_nodata}
    \end{figure}

In section~\ref{sec:comparisons-data}, we showed the predictions from \HEJ and
at NLO compared to LHC data at 8 and 13~TeV.  In order to make a realistic
comparison, we have added the ``HX'' component from the experimental papers.
Here, we include a few examples where the difference in shape resulting from
the all-order QCD treatment in \HEJ can be more clearly seen by studying only
the QCD component.  Figure~\ref{fig:8TeV_2j_nodata} shows this for two
distributions (originally shown in figures~\ref{fig:8TeV_dyjj} and
\ref{fig:8TeV_pTj3}).  Here we can see that the \HEJ predictions are strongly
suppressed compared to NLO as rapidity separation increases
(figure~\ref{fig:8TeV_dyjj_nodata}); however the transverse momentum spectrum is
harder for the third jet (figure~\ref{fig:8TeV_pTj3_nodata}).

